\documentclass[prd,aps,showpacs,floats,letterpaper,floatfix,superscriptaddress,twocolumn]{revtex4}

\usepackage{amssymb,amsmath}
\usepackage[dvips]{graphicx}

\newcommand{\hLN}{\hat{\mathbf{L}}_{N}}
\newcommand{\hL}{\hat{\mathbf{L}}}
\newcommand{\hS}{\hat{\mathbf{S}}}

\newcommand{\hJ}{\hat{\mathbf{J}}}
\newcommand{\vx}{\mathbf{x}}
\newcommand{\vv}{\mathbf{v}}

\newcommand{\vJ}{\mathbf{J}}
\newcommand{\vS}{\mathbf{S}}
\newcommand{\vL}{\mathbf{L}}
\newcommand{\glossaryitem}[1]{\emph{#1}:}

\newcommand{\beq}{\begin{equation}}
\newcommand{\eeq}{\end{equation}}
\newcommand{\bea}{\begin{eqnarray}}
\newcommand{\eea}{\end{eqnarray}}
\newcommand{\ba}{\begin{array}}
\newcommand{\ea}{\end{array}}

\newlength{\sizeonefig}
\newlength{\sizetwofig}
\begin{document}

\title{A quasi-physical family of gravity-wave templates for precessing 
binaries of spinning compact objects: II. Application to double-spin 
precessing binaries}

\author{Alessandra Buonanno}

\affiliation{Groupe de Gravitation et Cosmologie (GReCO), 
Institut d'Astrophysique de Paris (CNRS), 
98$^{\rm bis}$ Boulevard Arago, 75014 Paris, France}

\author{Yanbei Chen}

\affiliation{Theoretical Astrophysics and Relativity, California
Institute of Technology, Pasadena, CA 91125}

\author{Yi Pan}

\affiliation{Theoretical Astrophysics and Relativity, California
Institute of Technology, Pasadena, CA 91125}

\author{Michele Vallisneri}

\affiliation{Jet Propulsion Laboratory,
California Institute of Technology, Pasadena, CA 91109}

\begin{abstract}
The gravitational waveforms emitted during the adiabatic inspiral of
precessing binaries with two spinning compact bodies of comparable
masses, evaluated within the post-Newtonian approximation, can be
reproduced rather accurately by the waveforms obtained by setting one
of the two spins to zero, at least for the purpose of detection by
ground-based gravitational-wave interferometers. Here we propose to
use this quasi-physical family of single-spin templates to search for
the signals emitted by double-spin precessing binaries, and we find
that its signal-matching performance is satisfactory for source masses
$(m_1,m_2) \in [3,15]M_\odot \times [3,15]M_\odot$. For this mass
range, using the LIGO-I design sensitivity, we estimate that the
number of templates required to yield a minimum match of 0.97 is $\sim
320,000$. We discuss also the accuracy to which the single-spin
template family can be used to estimate the parameters of the original
double-spin precessing binaries.
\end{abstract}

\date{May 11, 2004}


\maketitle

\section{Introduction}
\label{sec1}

An international network of long-baseline laser-interferometric
gravitational-wave detectors, consisting of the Laser-Interferometer
Gravitational-wave Observatory (LIGO) \cite{LIGO}, of VIRGO
\cite{VIRGO}, of GEO\,600 \cite{GEO} and of TAMA\,300 \cite{TAMA}, has
by now begun science operations. VIRGO is in its commissioning phase,
while LIGO has already completed three science runs (S1 on
August--September 2002~\cite{S1}, S2 on February--April 2003, and S3
on October 2003--January 2004; S1 and S3 were in coincidence with
GEO\,600) with increasing sensitivity and stability. The analysis of
S1 data has been completed, yielding new upper limits on event rates
for various classes of astrophysical sources \cite{S1ULs}; the data
from S2 and S3 are still being analyzed. LIGO is expected to reach its
full design sensitivity in 2005.

Compact binaries consisting of black holes (BHs) and neutron stars
(NSs) are among the most promising \cite{burgay} and best understood sources for
such gravitational-wave (GW) interferometers, which can observe the
waves emitted by the binaries in the adiabatic-inspiral regime, where
post-Newtonian (PN) calculations
\cite{2PN,KWW,K,DJS,DJSd,BFIJ,IF,BDE,DIS3,bcv1} are appropriate to describe
the orbital dynamics and predict the gravitational
waveforms.

Very little is known about the statistical distributions of BH spins
in compact binaries: the spins could be large, and they need not be
aligned with the orbital angular momentum.
When this is the case, spin-orbit and spin-spin
interactions can cause the rapid precession of the orbital plane of
the binary, and thus significant modulations of the emitted GWs, as it
was shown by Apostolatos, Cutler, Sussman and Thorne (ACST)
\cite{ACST94}, and later by Apostolatos \cite{apostolatos0}. These
modulational effects should be included in the theoretical waveform
models (templates) used in matched-filtering GW searches. However,
using template banks parametrized by all the relevant physical
parameters (the masses, the spins, the angles that describe the
relative orientations of detector and binary, and the direction of
propagation of GWs to the detector) would make such searches extremely
computationally intensive.

One possibility to reduce the computational cost is the adoption of
smaller \emph{detection template families} (DTF), which capture the
essential features of the true waveforms, but depend on a smaller
number of parameters. A DTF for precessing binaries was first proposed
by Apostolatos~\cite{apostolatos1,apostolatos2}, building on the
analysis of precessional dynamics of
Refs.~\cite{ACST94,apostolatos0}. However, according to Apostolatos'
own estimates and to Grandcl\'ement, Kalogera and Vecchio's later
tests~\cite{GKV}, the computational resources required by the
Apostolatos DTF are still prohibitive, and its signal-matching
performance is unsatisfactory. The latter is improved in a modified
version of the DTF \cite{GK,Gpc}, which adds $\delta$-like
\emph{spikes} in the waveform phase.

Buonanno, Chen and Vallisneri \cite[henceforth BCV2]{bcv2}
investigated the dynamics of precession further, and proposed a new
convention to write the dominant quadrupolar contribution to GW
emission. In this convention, the oscillatory effects of precession
are isolated in the evolution of the GW polarization tensors, which
are combined with the detector's \emph{antenna patterns} to yield its
response. As a result, the response can be written as the product of a
carrier signal and a modulational correction, which can be handled
using an extension of Apostolatos' treatment of precessional
effects. BCV2 cast these waveforms into a mathematical structure that
allows searching automatically and economically over all the
parameters related to precessional modulations, except for a single
parameter that describes the timescale of modulation.  The BCV2 DTF
has reasonable computational requirements and good signal-matching
capabilities. However, especially for binaries with high, comparable
masses, it has the shortcoming that a large number of unphysical
waveforms are automatically included in GW searches (albeit at no
extra computational cost), increasing the probability of false alarms
triggered by noise.  [This shortcoming is unfortunately but
unavoidably characteristic of DTFs, which replace a description in
terms of physical source parameters by one in terms of
phenomenological signal parameters.]

This paper is the second in a series (begun with Ref.\ \cite{pbcv1}) written to investigate the possibility of
searching for precessing binaries using a \emph{physical} family of
signal templates (a PTF) computed from the PN equations of
motion. Although at first sight the number of physical parameters
necessary to describe a waveform is large, we were able to reduce the
effective dimensionality of the template family using the insight
developed in the construction of DTFs. As mentioned above, BCV2
\cite[Sec.\ VI~D]{bcv2} established that it is possible to search
easily over most of the parameters related to the kinematics of
precession (such as the orientation of the detector and of the binary
as a whole, the direction of GW propagation, and the initial orbital
phase). In effect, these \emph{extrinsic} parameters can be
incorporated in the detection statistic, while single ``templates''
\cite{note01} remain functions only of the masses of the binary components, of the
magnitudes of their spins, and of the relative angles between the
spins and the orbital angular momentum at a fiducial frequency. Under
the assumption of circular adiabatic inspiral \cite{circularadiabatic}, seven such
\emph{intrinsic} parameters are needed for a generic binary where both
spins are important (henceforth, a \emph{double-spin binary}); four
are needed for a binary where only one body has significant spin
(henceforth, a \emph{single-spin binary}).  (See Sec.\ \ref{secMF} for
the distinction between extrinsic and intrinsic parameters.)

In Ref.\ \cite{pbcv1}, we demonstrated the feasibility of a PTF
search for single-spin binaries: we described a two-stage algorithm to
search over the extrinsic parameters (the first stage emphasizes
computational efficiency, but retains some unphysical waveforms; the
second stage, applied only to first-stage triggers, restricts the
possible search outcomes to physical configurations). Using this
algorithm, we tested a four-parameter PTF for binaries with $(m_1,m_2)
\in [7M_\odot,12M_\odot] \times [1M_\odot,3M_\odot]$, where the
assumption of a single significant spin is justified. We found that
$\sim 76,000$ templates are required for a search in this mass range,
for a minimal match of 0.97 (see Sec.\ \ref{secMF} for a definition of
minimal match); under the assumption of Gaussian, stationary noise, we
also found that the detection threshold required for a given
false-alarm probability is lower in the PTF search than in a DTF
search with the same number of intrinsic parameters.

In this paper we examine PTF searches for the more general class of
double-spin binaries. Although in this case we have seven intrinsic
parameters, they are not all essential in determining the
waveforms. This is strictly true in two limits. First, as it was
realized by ACST~\cite{ACST94}, if the two binary masses are equal,
and if spin-spin interactions are ignored, the same orbital evolution
can be replicated by giving the total spin to one of the
objects. Indeed, for the mass ranges of interest to ground-based
interferometers, spin-spin effects contribute mildly to the binding
energy and to the PN GW flux, even close to the last stable
orbit. Second, if the mass ratio $\eta = m_1 m_2/(m_1+m_2)^2$ is very
low (as it was assumed in Ref.\ \cite{pbcv1}), the spin of the less massive
object can be ignored. In addition, as investigated by BCV2 (and less
systematically by Kidder~\cite{K}), the dynamics of double-spin
binaries with generic mass ratios show features similar to those
described by ACST for single-spin binaries.

These arguments have led us to conjecture that single-spin waveforms may always be sufficient to approximate double-spin waveforms, at least for the purpose of GW searches with ground-based interferometers. Since the single-spin parameters that best reproduce a given double-spin signal might not be in the physical range for a true single-spin binary (for instance, because the spin of one of two objects must do the work of two, it might have to exceed the maximal spin allowed for BHs), the single-spin family should be called \emph{quasi-physical}, but we shall continue to use ``PTF'' loosely. In the rest of this paper, we present evidence that our conjecture is correct for the mass range $(m_1,m_2) \in [3M_\odot,15M_\odot] \times [3M_\odot,15M_\odot]$, and we examine the computational requirements and the parameter-estimation performance of a single-spin PTF search for double-spin binaries in this mass range.

This paper is organized as follows. In Sec.~II, we provide a short glossary 
for the notions and quantities of matched-filtering GW searches (some standard, some developed in Ref.\ \cite{pbcv1}) that are needed later. In Sec.~III A, we review the adiabatic PN dynamics of double-spin binaries; in Sec.~III B, we describe our family of quasi-physical single-spin templates, and we evaluate their signal-matching capabilities against double-spin binaries with maximal spins (where precessional effects are expected to be strongest); in Sec.~III C, we study the robustness of adiabatic PN waveforms for binaries with high, comparable masses; in Sec.~III D, we discuss some features of double-spin binary dynamics that help to explain the good signal-matching performance of single-spin templates. In Sec.~IV, we estimate the number of templates required in a single-spin PTF search in our mass range of interest.
In Sec.~V, we investigate the extraction of the physical parameters of the double-spin binary using single-spin templates. Last, in Sec.~VI we summarize our main conclusions.

\section{A glossary of matched-filtering GW detection}
\label{secMF}

In this paper we adopt the standard formalism of matched-filtering GW
detection, as summarized in Ref.\ \cite{bcv1} (which includes an extensive bibliography), and as extended in Ref.\ \cite{pbcv1} to a
special treatment of extrinsic and intrinsic parameters. Here we
provide a glossary of the notions used in this paper, with pointers to
their definitions in Refs.\ \cite{bcv1,pbcv1}.

\glossaryitem{Templates $h(\lambda^A)$} Theoretical models of GW
signals, parametrized by one or more \emph{template parameters}
$\lambda^A$. A \emph{continuous template family} $\{h(\lambda^A)\}$
defines a smooth submanifold in signal space.

\glossaryitem{Noise inner product $\langle f, g \rangle$} Eq.\ (1) of
Ref.\ \cite{bcv1}. A measure of the closeness of two signal, as given
by a correlation product weighted by the power spectral density of
noise; throughout this paper we adopt the LIGO-I one-sided noise power spectral density $S_n$ given by Eq. (28) of Ref.\ \cite{bcv1}. A \emph{normalized} template $\hat{h}(\lambda^A)$ has $\langle \hat{h}(\lambda^A), \hat{h}(\lambda^A) \rangle = 1$.

\glossaryitem{Match} Inner product of two normalized signals. The
\emph{mismatch} is one minus the match.

\glossaryitem{Overlap $\rho(s,h(\lambda^A))$} Inner product of a
signal $s$ with the normalized template $\hat{h}(\lambda^A)$.

\glossaryitem{Detection statistic} Figure of merit compared with a
\emph{detection threshold} to decide whether the signal modeled by a
template $h(\lambda^A)$ is present in the detector output $o$. For
Gaussian, stationary noise, the overlap $\rho(o,h(\lambda^A))$ is the
\emph{optimal} statistic that minimizes the probability of false
dismissal for a given probability of false alarm (set by the detection
threshold). In this context $\rho$ is also known as the
\emph{signal-to-noise ratio} (S/N) of the detector output after
filtering by the template $\hat{h}(\lambda^A)$. The corresponding
detection statistic for an entire template family $\{h(\lambda^A)\}$
is the maximized overlap $\max_{\lambda^A} \rho(o,h(\lambda^A))$.

\glossaryitem{Fitting factor $\mathrm{FF}$ {\rm \cite{ffcits}}} Eq.\
(20) of Ref.\ \cite{bcv1}. Match between a template in a \emph{target}
family (representing actual physical signals) and a template in a
\emph{search} family, \emph{maximized over all the parameters of the
search family}. The FF (a value between 0 and 1) characterizes the
effectualness of the search family in reproducing signals modeled by
the target family: using an imperfect family means that only a
fraction FF of the available S/N is recovered, reducing the number of
true events that pass the detection threshold. The maximized match
induces a (many-to-one) map between the space of target parameters and
the space of search parameters (see Sec.~V on the
systematic errors in parameter estimation induced by this map).

\glossaryitem{Extrinsic $(\Theta^\mu)$ and intrinsic $(X^i)$ template
parameters} Extrinsic parameters are those over which $\rho$ can be
maximized efficiently, without recomputing full search templates for
each set of extrinsic parameters under consideration (but perhaps
using a small number of subtemplates) \cite{note02}. By contrast,
maximizing $\rho$ over the intrinsic parameters requires computing a
full template for each different set of intrinsic parameters in the
search range. Searches for signals modeled by a template family
$\{h(\lambda^A) \equiv h(X^i,\Theta^\mu)\}$ are usually implemented by
obtaining $\max_{\Theta^\mu} \rho$ for each template in a discrete bank
$\{h(X^i_{(k)},\Theta^\mu)\}$, laid down only along the
intrinsic-parameter directions.

\glossaryitem{Minimum match MM and mismatch metric {\rm
\cite{metriccits}}} Eqs.\ (21)--(24) of Ref.\ \cite{bcv1}. The spacing
of discrete search banks is chosen so that at most a fraction MM is
lost from the S/N that would be obtained with a continuous search bank; the corresponding loss in detection rate, for the same detection threshold, is a fraction $\mathrm{MM}^3$. The choice of the spacing is helped
by considering the (full) \emph{mismatch metric} $g_{AB}$, which
serves as a local quadratic expansion of the mismatch over all the
search parameters. An approximation to the number of templates needed to achieve a given MM, computed using the metric, is given by Eq.\ (25) of Ref.\ \cite{bcv1}.

\glossaryitem{Projected metric} Eqs.\ (65) and (72) of Ref.\ \cite{pbcv1}.  Given that the search template bank has no extension along the extrinsic-parameter
directions, it is useful to consider a projected metric
$g^\mathrm{proj}_{ij}$ that approximates the mismatch
\emph{already minimized over the extrinsic search parameters}. This
$g^\mathrm{proj}_{ij}$ is still a function of the intrinsic and
extrinsic \emph{target} parameters.  The \emph{average projected
metric} $\overline{g^\mathrm{proj}_{ij}}$ (Eq.\ (75) of Ref.\
\cite{pbcv1}) is a weighted average over the extrinsic target parameters, which can be used to estimate the number of templates needed to achieve a given
reduction in detection rate, for a uniform distribution of target
extrinsic parameters (this reduction is proportional to
$\overline{\mathrm{MM}}^3$, where $\overline{\mathrm{MM}}$ is the
\emph{average minimum match}).

\glossaryitem{Reduced search parameter space and reduction curves {\rm
\cite{redcits,pbcv1}}} It can happen that the variety of waveforms
spanned by an $n$-dimensional (search) template family is approximated
with very high FF by an $(n-k)$-dimensional subset of the family
(a \emph{reduced family}). This circumstance is signaled
locally in the mismatch metric by the presence of $k$ \emph{quasi-null
directions} (i.e., eigenvectors with very small eigenvalues). The
integral curves of these directions (the \emph{reduction curves})
correspond to sets of templates with very high match within the set,
and map a reduced family into another. In this case, it is advantageous to derive the discrete search bank from a reduced family: ideally, one would reparametrize the full family using $k$ parameters that run along the reduction curves, and then discard those parameters before laying down templates. See Ref.\ \cite{pbcv1} for a thorough discussion.

\section{Single-spin template family to match double-spin precessing binaries}
\label{sec2}

This section contains the main results of this paper.  In Sec.\
\ref{sec2.1} we describe the PN equations for the circular adiabatic
inspiral of a double-spin binary; this \emph{target model} is used
throughout this paper to represent physical signals. In Sec.\
\ref{sec2.2} we describe our proposed single-spin \emph{search
template family}, and we evaluate its effectualness (which is
excellent) in approximating the target waveforms. In Sec.\
\ref{sec2.3} we compare single-spin signals obtained at different PN
orders, to argue that the circular adiabatic model of inspirals used
in this paper gives robust predictions for the actual physical
waveforms. Last, in Sec.\ \ref{sec2.4} we study the precessional
dynamics of double-spin binaries to understand which of its features
can be represented accurately by single-spin systems, and which
cannot.

\subsection{Target model: double-spin precessing binaries}
\label{sec2.1}

Post-Newtonian calculations provide the following set of equations
describing the adiabatic evolution of double-spin precessing binaries.
The first derivative of the orbital (angular) frequency, up to 3.5PN
order \cite{note04} reads 
\cite{2PN,KWW,K,DJS,DJSd,BFIJ,IF,BDE,bcv2}:
\begin{widetext}
\beq
\frac{\dot{\omega}}{\omega^2} = \frac{96}{5}\,\eta\,(M\omega)^{5/3}
\left (1 + {\cal P}_{1\mathrm{PN}} + {\cal P}_{1.5\mathrm{PN}} + {\cal P}_{2\mathrm{PN}}
+ {\cal P}_{2.5\mathrm{PN}} + {\cal P}_{3\mathrm{PN}} + {\cal P}_{3.5\mathrm{PN}} \right ) \,,
\label{omegadot}
\eeq
where
\bea
{\cal P}_{1\mathrm{PN}} &=& -\frac{743+924\,\eta}{336}\,(M\omega)^{2/3} \,, \label{omegadotSTpn1} \\
{\cal P}_{1.5\mathrm{PN}} &=& -\Bigg
(\frac{1}{12}\sum_{i=1,2}\bigg[\chi_i\left(\hL_N\cdot\hS_i\right)\left(113\frac{m_i^2}{M^2}+75\eta\right)\bigg]
-4\pi\Bigg)(M\omega) \,,\\
\label{omegadotSTpn15}
{\cal P}_{2\mathrm{PN}} &=& \Bigg \{
\Bigg (\frac{34\,103}{18\,144}+\frac{13\,661}{2\,016}\,\eta+\frac{59}{18}\,\eta^2
\Bigg )-\frac{1}{48}\, \eta\,\chi_1\chi_2\left[247\,(\hS_1\cdot\hS_2)-721\,(\hL_N\cdot\hS_1)(\hL_N\cdot\hS_2)\right]
\Bigg \}\,(M\omega)^{4/3}\,, \\
\label{omegadotSTpn2}
{\cal P}_{2.5\mathrm{PN}} &=& -\frac{1}{672}\,(4\,159 +15\,876\,\eta)\,\pi\,(M\omega)^{5/3}\,, \\
\label{omegadotSTpn25}
{\cal P}_{3\mathrm{PN}} &=& \Bigg[
\left(\frac{16\,447\,322\,263}{139\,708\,800}-\frac{1\,712}{105}\,\gamma_E+\frac{16}{3}\pi^2\right)+
\left(-\frac{273\,811\,877}{1\,088\,640}+\frac{451}{48}\pi^2-\frac{88}{3}\hat\theta
\right)\eta \nonumber \\
&&+\frac{541}{896}\,\eta^2-\frac{5\,605}{2\,592}\,\eta^3
-\frac{856}{105}\log\left[16(M\omega)^{2/3}\right] \Bigg] (M\omega)^2\,,\\
\label{omegadotSTpn3}
{\cal P}_{3.5\mathrm{PN}} &=& \Bigg (
-\frac{4\,415}{4\,032}+\frac{358\,675}{6\,048}\,\eta+\frac{91\,495}{1\,512}\,\eta^2
\Bigg )\,\pi\,(M\omega)^{7/3}\,.
\label{omegadotSTpn35}
\eea
\end{widetext}
Here, $m_1$ and $m_2$ are the masses of the two bodies, with $m_1 \geq
m_2$; $M=m_1+m_2$ is the total mass, and $\eta = m_1 m_2 / M^2$ is the
symmetric mass ratio; $\vL_N = \mu \, \vx \times \vv$ (with $\mu = m_1 m_2/M$)
is the Newtonian angular momentum (with $\vx$ and $\vv$ the two-body center-of-mass
radial separation and relative velocity), and $\hL_N = \vL_N /
|\vL_N|$; $\vS_1 =\chi_1\,m_1^2\,\hS_1$ and $\vS_2
=\chi_2\,m_2^2\,\hS_2$ are the spins of the two bodies (with
$\hS_{1,2}$ unit vectors, and $0 < \chi_{1,2} < 1$ for BHs);
$\gamma_E=0.577\ldots$ is Euler's constant; last, $\hat{\theta}=
\theta +1987/1320 + 7\omega_s/11$ (with $\theta$ an unknown arbitrary
parameter that enters the GW flux at 3PN order~\cite{BFIJ},
and $\omega_s = 0$~\cite{DJSd,IF,BDE}).
(Note for v3 of this paper on gr-qc: Eqs.\ (5) and (7) are now revised as per Ref.\ \cite{errata};
the parameter $\hat{\theta}$ has been determined to be 1039/4620 \cite{thetapar}.)
 
In Eq.~\eqref{omegadot} we did not include the
quadrupole-monopole terms~[\onlinecite{QM}], because we have already
shown (see Sec.\ III E of Ref.~[\onlinecite{pbcv1}]) that those terms
do not significantly affect matches once these are maximized on binary
parameters. The precession equations for the spins read
[\onlinecite{K},\onlinecite{ACST94}]
\begin{widetext}
\bea
\label{S1dot}
\dot{\vS}_1&=&
\frac{(M\omega)^2}{2M}
\left\{ \eta\,(M\omega)^{-1/3}\left(4+3\frac{m_2}{m_1}\right)\hL_N
+ \frac{1}{M^2}\,\left[\vS_2-3(\vS_2\cdot\hL_N)\hL_N\right]\right\}\times\vS_1
\,,\\
\label{S2dot}
\dot{\vS}_2&=&
\frac{(M\omega)^2}{2M}
\left\{ \eta\, (M\omega)^{-1/3}\left(4+3\frac{m_1}{m_2}\right)\hL_N
+ \frac{1}{M^2}\,\left[\vS_1-3(\vS_1\cdot\hL_N)\hL_N\right]\right\}\times\vS_2
\,,
\eea
and the precession equation for $\hL_N$ is [\onlinecite{K},\onlinecite{ACST94}]
\beq
\dot{\hL}_N=\frac{\omega^2}{2M}
\Bigg\{\left[\left(4+3\frac{m_2}{m_1}\right)\,\vS_1+
\left(4+3\frac{m_1}{m_2}\right)\,\vS_2\right]\times\hL_N
-\frac{3\,\omega^{1/3}}{\eta\,M^{5/3}}\left[(\vS_2\cdot\hL_N)\,\vS_1+
(\vS_1\cdot\hL_N)\,\vS_2\right]\times\hL_N \Bigg\}\,.
\label{Lhdot}
\eeq
\end{widetext}
We stop the adiabatic evolution when the binary reaches the minimum energy 
circular orbit (MECO) defined by Eqs.~(11)--(13) of Ref.~[\onlinecite{bcv2}]. 
The leading-order mass-quadrupole gravitational waveform can be obtained 
from Eqs.~(65)--(78) of Ref.~[\onlinecite{bcv2}]. Those equations 
for the gravitational waveform, together with Eqs.~(\ref{omegadot})--(\ref{Lhdot}), define our target model for precessing double-spin binaries.

Using the language of Ref.~\cite{bcv2}, precessing binaries of spinning BHs 
are described by the four {\it basic} (intrinsic) parameters $m_1$, $m_2$, $S_1 \equiv |\vS_1|$, $S_2 \equiv |\vS_2|$, by three {\it local} (intrinsic) parameters describing the relative orientation of the spins with respect to the angular momentum at a fiducial frequency (see Table I and Fig.\ 4 of Ref.~\cite{bcv2}), and by five {\it directional} (extrinsic) parameters describing the relative orientation of the binary and the detector (see Table I of Ref.~\cite{bcv2}). The waveforms depend also (if trivially) on two other extrinsic parameters, the initial phase $\Phi_0$ and the time of arrival $t_0$; depending on the context, we shall at times omit these when counting the number of extrinsic parameters.

\subsection{Search template family: single-spin binaries}
\label{sec2.2}

As discussed in Sec.~\ref{sec1}, the results of previous investigations~[\onlinecite{ACST94}, \onlinecite{CF94}, \onlinecite{bcv2}, \onlinecite{pbcv1}] suggest that the gravitational waveforms emitted by double-spin binaries with comparable component masses can be approximated (at least for the purpose of detection with ground-based interferometers) by waveforms computed by neglecting spin-spin effects, and by assigning the total spin of the binary to a single BH. Thus, in this paper we examine the detection performance of the single-spin search family obtained from Eqs.\ (\ref{omegadot})--(\ref{Lhdot}) by setting $\vS_2 = 0$. The simplified equations are 
\bea
\label{templ1}
\frac{\dot{\omega}}{\omega^2} &=&\frac{96}{5}\,\eta_{s}\,(M_{s}\, \omega)^{5/3}\,
\biggl[ 1 + \mbox{PN \,corrections} \\ \nonumber &-&
\frac{1}{12}\!\left(\hL_N\cdot\hS_{1s} \right)\!\chi_{1s}\!
\left(113\,\frac{m_{1s}^2}{M_{s}^2} +75 \eta_{s} \!\right)
\! (M_{s}\,\omega) \biggr] \,,\\
\label{templ2}
\dot{\mathbf{S}}_{1s} &=&
\frac{\eta_{s} (M_{s}\, \omega)^{5/3}}{2M_{s}}\,
\left (4+ 3 \frac{m_{2s}}{m_{1s}} \right )\,
\hL_N \times \mathbf{S}_{1s}\,,\\
\label{templ3}
\dot{\hL}_{N}&=&
\frac{\omega^{2}}{2 M_{s}}\,\left ( 4 +3 \frac{m_{2s}}{m_{1s}}\right )\mathbf{S}_{1s} \times \hL_N\,,
\eea
where $M_{s} = m_{1s} + m_{2s}$, $\eta_{s} = 
m_{1s}\,m_{2s}/M^2_{s}$ and $\vS_{1s} = \chi_{1s}\, 
m_{1s}^2\,\hS_{1s}$. The ``\textit{s}'' subscript stands for  \emph{search} parameters. In Eq.~(\ref{templ1}), ``PN corrections'' denotes the terms in Eqs.~(\ref{omegadotSTpn1})--(\ref{omegadotSTpn35}) that do not depend on the spins. The leading-order mass-quadrupole gravitational waveform is given by Eqs.~(65)--(78) and (11)--(13) of Ref.~[\onlinecite{bcv2}], after setting the spin of the lighter body to zero. This completes the definition of our single-spin search template family, which is parametrized by the four intrinsic parameters $M_{s}$, $\eta_{s}$, $\chi_{1s}$, and $\kappa_{1s} = \hS_{1s}\cdot \hL_N$, and by five extrinsic parameters that describe the relative orientation of the detector and the binary (see Sec.\ III C of Ref.\ \cite{pbcv1}). The maximization of the overlap with respect to the extrinsic parameters can be performed semi-algebraically, in two steps, as described in Sec.\ IV of Ref.~\cite{pbcv1}.
\begin{table*}
\begin{center}

{\begin {tabular}{r||r|r|r|r|r|r|r|r}
& \multicolumn{2}{c|} {$(3+3)M_\odot$} & $(6+3)M_\odot$ &
 \multicolumn{2}{c|} {$(6+6)M_\odot$} & $(9+3)M_\odot$ &
 \multicolumn{2}{c} {$(12+3)M_\odot$} \\
templates: & with spin & nospin & with spin & with spin & no spin & with
 spin & with spin & equal-spin target \\
\hline\hline
FF $\ge 0.99$ & 95\% & 31\% & 74\% & 98\% & 59\% & 90\% & 95\% & 84\% \\
FF $<0.99$ & 5\% & 69\% & 26\% & 2\% & 41\% & 10\% & 5\% & 16\% \\ 
FF $<0.95$ & 0\% & 38\% & 3\% & 0\% & 25\% & 1\% & 0\% & 0\% \\ 
lowest FF & 0.9085 & 0.7042 & 0.9119 & 0.7250 & 0.6391 & 0.8945 & 0.9734 & 0.9684 \\ 
$\overline{\rm FF}$ & $\ge 0.989$ & $\ge 0.938$ & $\ge 0.986$ & $\ge
 0.987$ & $\ge 0.934$ & $\ge 0.989$ &
 $\ge 0.990$ & $\ge 0.990$ \\
\end {tabular}}

\vspace{0.5cm}

{\begin {tabular}{r||r|r|r|r|r|r}
& \multicolumn{2}{c|}{$(10+10)M_\odot$} &
 \multicolumn{2}{c|}{$(15+10)M_\odot$} &
 \multicolumn{2}{c}{$(15+15)M_\odot$} \\
templates: & with spin & nospin & with spin & no spin & with spin & no
 spin \\ 
\hline\hline
FF $\ge 0.99$ & 100\% & 29\% & 98\% & 22\% & 100\% & 30\% \\ 
FF $< 0.99$ & 0\% & 71\% & 2\% & 78\% & 0\% & 70\% \\ 
FF $<0.95$ & 0\% & 34\% & 0\% & 46\% & 0\% & 31\% \\ 
lowest FF & 0.9754 & 0.7142 & 0.9691 & 0.7138 & $\ge 0.99$ & 0.7546 \\ 
$\overline{\rm FF}$ & $\ge 0.990$ & $\ge 0.945$ & $\ge 0.990$ & $\ge 0.936$ & $\ge
 0.990$ & $\ge 0.957$ \\
\end {tabular}}
\end{center}
\caption{Summary of FFs between the single-spin search template family
and the double-spin target model. The numerical maximization procedure
is stopped whenever a FF $\ge 0.99$ is achieved. The upper table shows
results for lower-mass binaries ($M \leq 15 M_\odot$); the lower table
for higher-mass binaries ($M \geq 20 M_\odot$). In the first three
rows of each table, we list the percentage of systems yielding FFs $\ge
0.99$, $<0.99$ and $<0.95$, in a population of 100 target systems [500
for $(m_1+m_2) = (6+3)M_\odot$ and $(9+3)M_\odot$ binaries] with
maximal spins and random, uniform distributions of initial spin and
detector orientations (local parameters). In the fourth row we list
the lowest FFs found among the population; in the last row, we list
the average FFs [when a FF $\ge 0.99$, we use $0.99$ in computing the
average.]  The distribution of the FFs for selected mass
configurations is also histogrammed in Fig.\ \ref{myhist}. The target
and search waveforms are computed by starting the integration of the
equations of motion at an instantaneous GW frequency of is $60$ Hz and
$40$ Hz for upper and lower tables, respectively. For some mass configurations
we show also the FFs for nonspinning templates (i.e.,
single-spin templates where $\chi_{1s}$ was set to zero), and for
$(12+3)M_\odot$ binaries, we show the FFs for a target configuration with
$\chi_1=1/16$ and $\chi_2=1$ (i.e., the $S_2$ is maximal and $S_1 = S_2$).
\label{Tab1}}
\end{table*}
\begin{figure}
\begin{center}
\includegraphics[width=3.2in]{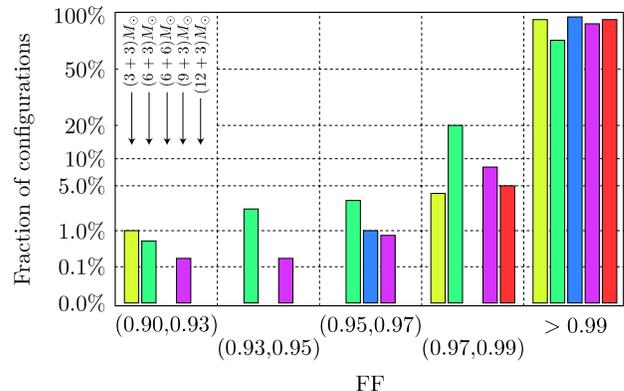}
\caption{\label{myhist} Distribution of FFs for lower-mass ($M \leq 15
M_\odot$) binary configurations. See the caption to Table \ref{Tab1}
for an explanation of how the FF distributions were obtained.}
\end{center}
\end{figure}

We note that the simplified Eqs.~\eqref{templ1}--\eqref{templ3} are
exactly equivalent to the full Eqs.\ \eqref{omegadot}--\eqref{Lhdot}
in the two limits mentioned in Sec.\ I: for equal masses, if spin-spin
effects are neglected; and for $m_1 \gg m_2$ (i.e., small $\eta$). To
test the effectualness of the single-spin search templates in matching
our target signals for binary configurations with both comparable and
unequal masses, we computed the FF (i.e., the match, maximized over the intrinsic and extrinsic search parameters; see Sec.\ II) for target binaries with
two maximal spins, and with masses $(m_1+m_2) = (3+3)M_\odot$,
$(6+3)M_\odot$, $(6+6)M_\odot$, $(9+3)M_\odot$, $(12+3)M_\odot$,
$(10+10)M_\odot$, $(15+10)M_\odot$, and $(15+15)M_\odot$. 
In the $(12+3)M_\odot$ case we also considered a target 
binary with $\chi_1 = 1/16$ and $\chi_2 = 1$, i.e. the two objects
possess equal magnitude of spins, while the less massive one is maximally
spinning. Search and target signals were always obtained at 2PN order,
and the computation
of the FF was repeated for 100 or 500 (for the lighter binaries)
randomly generated configurations of the target-signal local
parameters $\hLN$, $\hS_1$ and $\hS_2$, assuming uniform and
independent angular distributions. The directional parameters of the
target signals were fixed to arbitrary values without loss of
generality, since for the purpose of computing FFs they are degenerate
with respect to the local parameters (see Sec.\ VI A of
Ref.~\cite{bcv2}).

The results of our tests are listed in Table~\ref{Tab1}, and plotted
in Fig.~\ref{myhist}. For comparison, in Table~\ref{Tab1} we include
also some FFs computed for the nonspinning search templates obtained
by setting $\chi_{1s}$ to zero in Eqs.\ \eqref{templ1}--\eqref{templ3}.  Our numbers support our conjecture about the effectualness of the single-spin search family. More
specifically:
\begin{itemize}
\item Spin-spin effects are not important for higher-mass
binaries such as $(15+15)M_\odot$, $(15+10)M_\odot$ and $(10+10)M_\odot$, where FFs are consistently very high; however, spin-orbit effects cannot be neglected, as shown by the relatively low FFs achieved by the nonspinning search family.
\item For lower-mass binaries such as $(3+3)M_\odot$ and $(6+6)M_\odot$, FFs are also very high, with few exceptions: thus, although in these binaries spin-spin effects can accumulate over many GW cycles within the band of good detector sensitivity, they rarely become comparable to spin-orbit effects.
\item For low--mass-ratio binaries such as $(12+3)M_\odot$ ($\eta=0.16$), FFs are high, since the spin of the heavier object dominates the precessional dynamics. If we reduce the magnitude of $S_1$ so that $S_1 = S_2$, the resulting FFs become lower, because the dynamics deviates farther from both the single-spin and equal-mass limits.
\item The worst FFs are obtained for $(6+3)M_\odot$ and $(9+3)M_\odot$ binaries, which have rather low total masses, and intermediate mass ratios (thus, they sit halfway between the two single-spin equivalence limits). In this case, double-spin effects cannot be reproduced with accuracy by single-spin systems (in Sec.\ III D we shall examine in more detail what is happening there). Note however that this happens only for a limited number of angular configurations, so the average of the FF over the sampling is still very high.  
\end{itemize}
\begin{figure*}
\begin{center}
\includegraphics{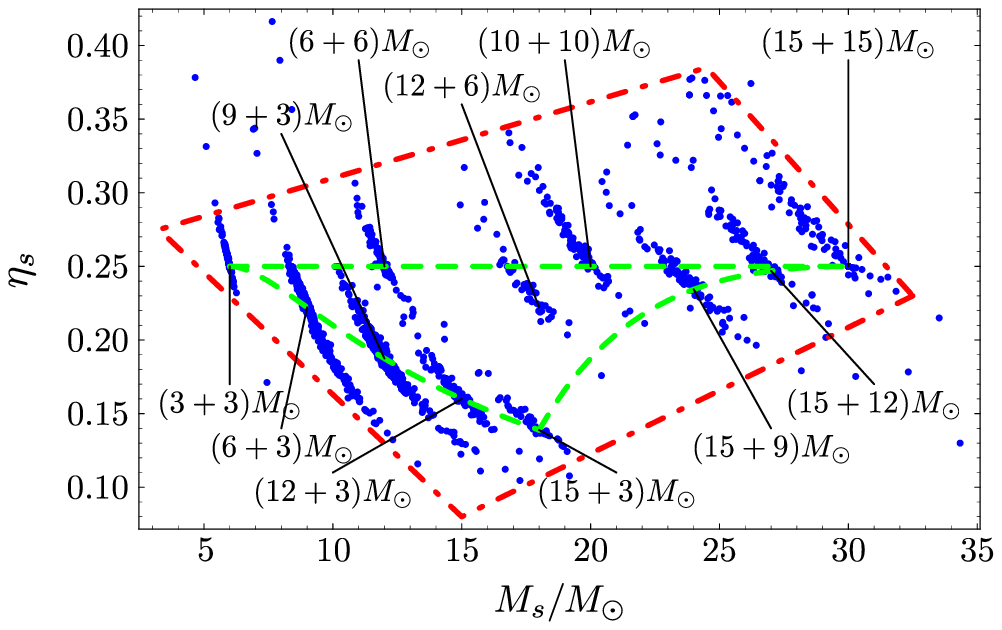} \\[0.5cm]
\includegraphics[width=3in]{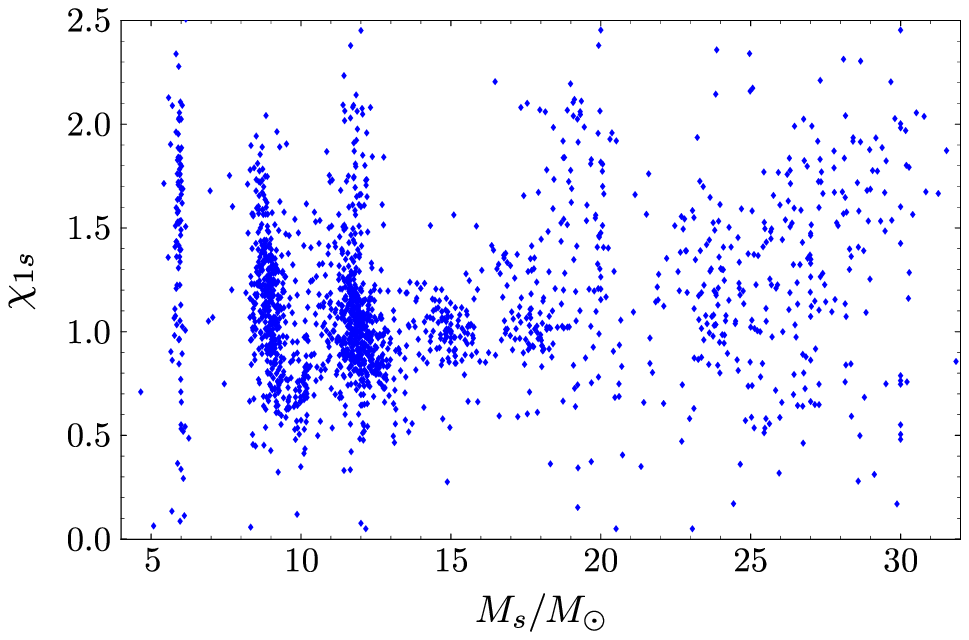} \hspace{0.5cm}
\includegraphics[width=3in]{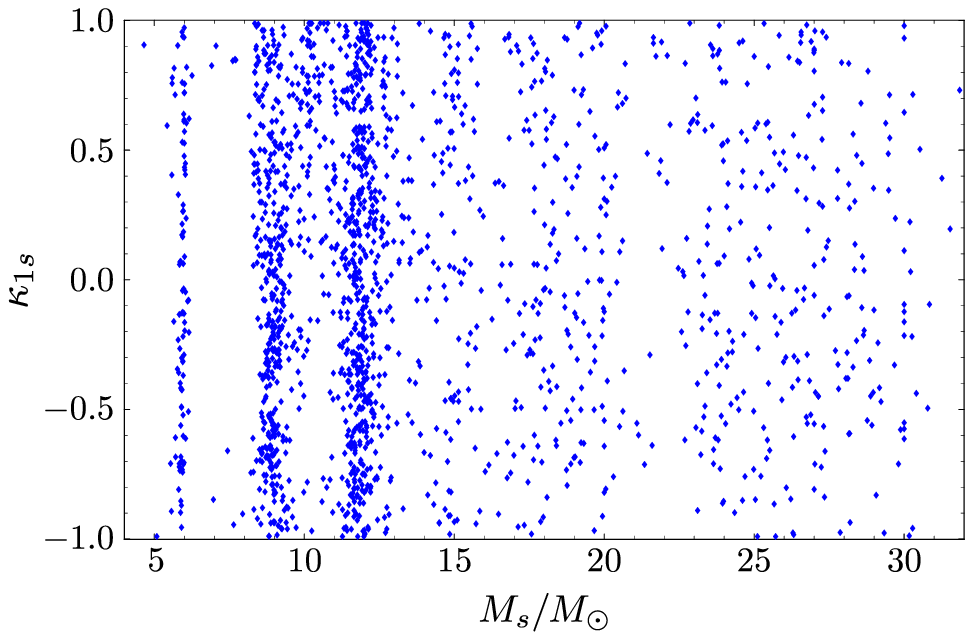}
\caption{Location in the (intrinsic) search parameter space
$(M_s,\eta_s,\chi_{1s},\kappa_{1s})$ of the best-fit templates for target
signals with $(m_1+m_2) = (3+3)M_\odot$, $(6+3)M_\odot$,
$(9+3)M_\odot$, $(12+3)M_\odot$, $(15+3)M_\odot$, $(6+6)M_\odot$,
$(12+6)M_\odot$, $(10+10)M_\odot$, $(15+10)M_\odot$, $(15+12)M_\odot$,
and $(15+15)M_\odot$, with maximal spins, and with random angular distributions of the initial
$\hL_N$, $\vS_1$, $\vS_2$. Dots are denser for the $(6+3)M_\odot$ and $(9+3)M_\odot$ configurations, for which more FF were computed.
In the $(M_s,\eta_s)$ scatter plot (on top), the dashed contour
encloses the region obtained by setting $M_s = M$ and
$\eta_s = \eta$, and by taking $(m_1,m_2) \in [3,15] M_\odot \times
[3,15] M_\odot$. The dotted and dashed line, drawn somewhat
arbitrarily, encloses a possible template bank boundary, used in Sec.\
\ref{seccounting} to estimate the number of templates necessary to
search for double-spin binaries in this mass range. The labels identify the search template clusters corresponding to each target mass configuration, and they are connected to the nominal projection point obtained by setting $M_s = M$ and $\eta_s = \eta$.
\label{fig:scatter}}
\end{center}
\end{figure*} 

The range of search-template parameters needed to yield the high FFs
discussed above extends beyond values that would be physical for a
real single-spin binary, with $\eta_{s}>0.25$ and $\chi_{1s} >
1$. This is to be expected: consider, for instance, that in the
equal-mass case the equivalence between the simplified and the full
equations implies values of $\chi_{1s}$ up to 2.  In
Fig.~\ref{fig:scatter} we show the parameters of the best-fit search
templates corresponding to target signals with the test masses
examined above [augmented by $(15+3)M_\odot$, $(12+6)M_\odot$, and
$(15+12)M_\odot$]. As shown in the top panel, the search-template
images of target signals with the same masses but different local
parameters are spread around the nominal $(M_s,\eta_s)$ values
(indicated by the end of the thin lines, and always enclosed within
the dashed contour). The uncertainty in target parameter estimation
induced by this spreading is discussed in Sec.\ \ref{sec4}. In the
same panel, the dotted-dashed line encloses the template-bank boundary
used in Sec.\ \ref{seccounting} to estimate the number of templates
necessary for a search of systems with masses $(m_1,m_2) \in [3,15]
M_\odot \times [3,15] M_\odot$. The bottom panels show the range
achieved by the search parameters $\chi_{1s}$ and $\kappa_{1s}$, which
is comparable to the range of the analogous target parameters,
$|\vS_{\rm tot}|/m_1^2$ and $\kappa_\mathrm{tot} \equiv \hS_{\rm tot}
\cdot \hL_{N}$.

\subsection{On the robustness of waveforms across PN orders}
\label{sec2.3}
\begin {table*} 
\begin {tabular}{l||rr|rr|rr|rr}
\multicolumn{1}{c||}{($N+k$,$N$)} & \multicolumn{8}{c}{$\left <{\rm
ST}_{N+k},{\rm ST}_{N} \right >$ for $(10+10)M_\odot$ binary, $M=
20M_\odot,\;\;\eta= 0.250$} \\ & \multicolumn{2}{c|}{$\kappa_1 = 0.9$}
& \multicolumn{2}{c|}{$\kappa_1 = 0.5$} &
\multicolumn{2}{c|}{$\kappa_1 = -0.5$} & \multicolumn{2}{c}{$\kappa_1
= -0.9$}\\ \hline\hline (1,0) & 0.3136 & (0.6688) [16.1,0.25,0.00,
0.00] & 0.3136 & (0.6688) & 0.3136 & (0.6688) [16.1,0.25,0.00,$-$0.00]
& 0.3136 & (0.6688) \\ (1.5,1) & 0.3123 & (0.5922) [23.7,0.25,0.00,
0.00] & 0.2676 & (0.5137) & 0.2306 & (0.4860) [29.2,0.25,0.00,$-$0.00]
& 0.2160 & (0.4543) \\ (2,1.5) & 0.7124 & (0.9823) [19.2,0.25,1.00,
0.99] & 0.7222 & (0.9877) & 0.8545 & (0.9886)
[19.3,0.25,1.00,$-$0.66] & 0.8601 & ($\ge$0.99) \\ (2.5,2) & 0.2851 &
(0.8702) [18.8,0.25,1.00, 0.91] & 0.3099 & (0.9206) & 0.4166 &
($\ge$0.99) [20.4,0.25,1.00,$\;\;\:$0.81] & 0.4682 & ($\ge$0.99) \\ (3,2) &
0.9604 & ($\ge$0.99) [20.2,0.25,0.88, 0.79] & 0.9743 & ($\ge$0.99) &
0.9915 & ($\ge$0.99) [19.7,0.24,0.88,$-$0.23] & 0.9805 & ($\ge$0.99)\\
(3,2.5) & 0.2848 & (0.9846) [18,8,0.25,1.00,-0.31] & 0.2898 &
($\ge$0.99) & 0.4027 & (0.9823) [20.4,0.25,1.00,$-$0.99] & 0.4634 &
(0.9740)\\ (3.5,3) & 0.9316 & ($\ge$0.99) [20.2,0.25,1.00, 0.90] &
0.9475 & ($\ge$0.99) & 0.9749 & ($\ge$0.99) [19.7,0.25,1.00,$-$0.99] &
0.9744 & ($\ge$0.99)
\end {tabular}

\vspace{0.5cm} 

\begin {tabular}{l||rr|rr|rr|rr}
\multicolumn{1}{c||}{($N+k$,$N$)} & \multicolumn{8}{c}{$\left <{\rm
ST}_{N+k},{\rm ST}_{N} \right >$ for $(15+10)M_\odot$ binary, $M =
25M_\odot, \;\;\eta=0.247$} \\ & \multicolumn{2}{c|}{$\kappa_1 = 0.9$}
& \multicolumn{2}{c|}{$\kappa_1 = 0.5$} &
\multicolumn{2}{c|}{$\kappa_1 = -0.5$} & \multicolumn{2}{c}{$\kappa_1
= -0.9$}\\ \hline\hline (1,0) & 0.3124 & (0.6030)
[19.1,0.25,0.00,0.00] & 0.3124 & (0.6030) & 0.3124 & (0.6030)
[19.1,0.25,0.00,$-$0.00] & 0.3124 & (0.6030) \\ (1.5,1) & 0.2784 &
(0.4994) [31.0,0.25,0.00,0.00] & 0.2732 & (0.4684) & 0.2466 & (0.3913)
[41.9,0.25,0.00,$-$0.00] & 0.1896 & (0.3491) \\ (2,1.5) & 0.5810 &
($\ge$0.99) [24.8,0.23,1.00,0.98] & 0.8038 & ($\ge$0.99) & 0.8644 &
($\ge$0.99) [25.4,0.24,1.00,$-$0.59] & 0.9067 & ($\ge$0.99) \\ (2.5,2)
& 0.2558 & (0.8525) [22.9,0.23,1.00,0.95] & 0.3296 & (0.9280) & 0.5022
& ($\ge$0.99) [25.5,0.24,1.00,$\;\;\:$0.80] & 0.5921 & ($\ge$0.99) \\
(3,2) & 0.9106 & ($\ge$0.99) [24.8,0.24,0.88,0.73] & 0.9392 &
($\ge$0.99) & 0.9858 & ($\ge$0.99) [25.1,0.23,0.89,$-$0.21] & 0.9650 &
($\ge$0.99)\\ (3,2.5) & 0.2520 & (0.9148) [27.0,0.25,1.00,0.85] &
0.2942 & ($\ge$0.99) & 0.4552 & ($\ge$0.99) [25.7,0.24,1.00,$-$0.59] &
0.5333 & ($\ge$0.99)\\ (3.5,3) & 0.9264 & ($\ge$0.99)
[24.9,0.24,1.00,0.90] & 0.9528 & ($\ge$0.99) & 0.9769 & ($\ge$0.99)
[25.3,0.24,1.00,$-$0.36] & 0.9839 & ($\ge$0.99)
\end {tabular}

\vspace{0.5cm}

\begin {tabular}{l||rr|rr|rr|rr}
\multicolumn{1}{c||}{($N+k$,$N$)} & \multicolumn{8}{c}{$\left <{\rm
ST}_{N+k},{\rm ST}_{N} \right >$ for $(15+15)M_\odot$ binary, $M =
30M_\odot, \;\;\eta=0.250$} \\ & \multicolumn{2}{c|}{$\kappa_1 = 0.9$}
& \multicolumn{2}{c|}{$\kappa_1 = 0.5$} &
\multicolumn{2}{c|}{$\kappa_1 = -0.5$} & \multicolumn{2}{c}{$\kappa_1
= -0.9$}\\ \hline\hline (1,0) & 0.2710 & (0.5158)
[22.5,0.25,0.00,$\;\;\:$0.00] & 0.2710 & (0.5158) & 0.2710 & (0.5158)
[22.5,0.25,0.00,$-$0.00] & 0.2710 & (0.5158) \\ (1.5,1) & 0.2694 &
(0.4050) [38.3,0.25,0.00,$\;\;\:$0.00] & 0.2145 & (0.3644) & 0.2435 &
(0.3155) [49.6,0.25,0.00,$-$0.00] & 0.1855 & (0.2797) \\ (2,1.5) &
0.7619 & ($\ge$0.99) [31.1,0.25,1.00,$\;\;\:$0.89] & 0.8613 &
($\ge$0.99) & 0.9018 & ($\ge$0.99) [30.6,0.25,1.00,$-$0.43] & 0.8946 &
($\ge$0.99) \\ (2.5,2) & 0.3403 & (0.9086)
[28.0,0.24,0.92,$\;\;\:$0.99] & 0.3856 & (0.9237) & 0.5372 &
($\ge$0.99) [30.4,0.25,1.00,$\;\;\:$0.95] & 0.6216 & ($\ge$0.99) \\
(3,2) & 0.9330 & ($\ge$0.99) [28.8,0.24,0.90,$\;\;\:$0.82] & 0.9360 &
($\ge$0.99) & 0.9641 & ($\ge$0.99) [30.3,0.24,0.88,$\;\;\:$0.17] &
0.9612 & ($\ge$0.99)\\ (3,2.5) & 0.2926 & ($\ge$0.99)
[28.1,0.25,1.00,$-$0.58] & 0.3549 & ($\ge$0.99) & 0.4893 & ($\ge$0.99)
[31.0,0.25,1.00,$-$0.67] & 0.5498 & ($\ge$0.99)\\ (3.5,3) & 0.9265 &
($\ge$0.99) [30.4,0.25,0.90,$\;\;\:$0.71] & 0.9456 & ($\ge$0.99) &
0.9814 & ($\ge$0.99) [29.1,0.25,1.00,$-$0.30] & 0.9831 & ($\ge$0.99)
\end {tabular}

\caption{\label{CauchyST123} Test of robustness of the PN adiabatic
waveforms $\mathrm{ST}_N$ (defined in Sec.\ \ref{sec2.1}) across PN
orders, for $(m_1+m_2) = (10+10)M_{\odot}$, $(15+10)M_{\odot}$ and
$(15+15)M_{\odot}$. We set $\chi_1 = 1$, $\chi_2 = 0$, and $\kappa_1 =
-0.9, -0.5, 0.5$, and 0.9. The matches quoted at the beginning of each
column are maximized only with respect to the extrinsic parameters
$t_0$ and $\Phi_0$. In parentheses, ``(...)'', we give the matches
maximized over all the parameters of the lower-order family (i.e., the
fitting factors FF for the target family $\mathrm{ST}_{N+k}$ as
matched by the search family $\mathrm{ST}_N$, evaluated at the
$\mathrm{ST}_{N+k}$ intrinsic parameters indicated). In brackets,
``[...]'', we give the parameters $M$, $\eta$, $\chi_1$, and $\kappa_1$ (or $M$ and $\eta$ at 1PN and 1.5PN orders when spin terms are absent) at which the FF is achieved.  The detector is set
perpendicular to the initial orbital plane, and at 3PN and 3.5PN order
we set $\hat\theta=0$; in all cases the integration of the equations
of motion starts at an instantaneous GW frequency of $40$ Hz. 
[See Refs.~\cite{bcv1,bcv2} for a discussion of why for some mass combinations 
the 2.5PN model differs so much from the other orders.]}
\end {table*}

In the previous section we have established that single-spin waveforms
are good approximations for double-spin waveforms, at least within the
mass range under consideration. However, whether double-spin waveforms
are representative of actual physical signals is an entirely different
question, which hinges on the validity of the circular adiabatic
approximation, but also on the robustness of the waveforms under
change of PN order: if the waveforms change substantially with
increasing order, we should suspect that the description of the
physics is incomplete without higher-order terms yet to be computed.

Studying the robustness of double-spin binaries is technically
difficult, since it means computing the FF between two template
families (of different PN order) with seven intrinsic parameters. This
entails the delicate numerical maximization of a seven-parameter
function whose evaluation is relatively costly. Instead, we choose to
perform our study on single-spin waveforms, and then argue that the
results should transfer to double-spin waveforms because the search
and target families are close. Thus, our study is very similar to the
Cauchy convergence test of Ref.~\cite{pbcv1}, except for the
choice of masses: here we focus on binaries at the higher-mass end of
our range, since these systems are expected to have stronger
higher-order PN effects within the frequency band of good
interferometer sensitivity.

For $(m_1+m_2) = (10+10) M_\odot$, $(15+10) M_\odot$, and $(15+15)
M_\odot$, we list in Table~\ref{CauchyST123} the matches across PN
orders, maximized only on $t_0$ and $\Phi_0$; the numbers in
parentheses, ``(...)'', give the FF for the higher-order family as
matched by the lower-order family, and the numbers in brackets,
``[...]'', give the intrinsic parameters where the FF is attained. The
tests are performed for $\chi_{1s} = 1$, for different values of
$\kappa_{1s}$, and for a GW detector in a direction (with respect to the binary) perpendicular to normal vector of the initial
orbital plane, which should be representative of the
generic effects of precession. The high FFs obtained between the 2PN
and the higher-order families suggest that the 2PN model is already
representative of the variety of waveforms expected from actual
sources; on the other hand, the lower direct matches (and the biased
values of search parameters at the FF) suggest that the family of the
highest available order should be used for source parameter
estimation. It would be worthwhile to evaluate the FF between the double-spin (and indeed,
single-spin) adiabatic model, and nonadiabatic models based on
resummed PN equations~\cite{Pade-EOB,DIS3,bcv1,DIJS}, especially when
these predict the end of the inspirals within the band of good
interferometer sensitivity.

\subsection{Some features of the dynamics of double-spin binaries}
\label{sec2.4}
\begin{figure}
\begin{center}
\vspace{0.2cm}
\includegraphics[width=3.1in]{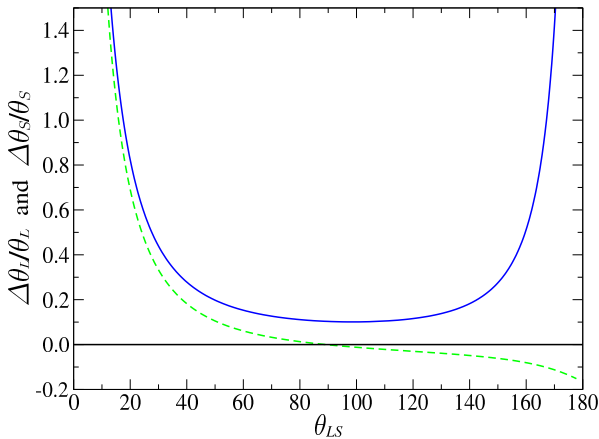}
\caption{\label{dopenang}
Relative change of the opening angles as function of $\theta_{LS}$ for 
a $(6+3)M_{\odot}$ binary, with $\chi_{\rm tot}=0.4$, $|\vL|=|\vL_N| = 
\eta M^{5/3}\,\omega^{-1/3}$, and $\omega = 2 \pi \times 30$ Hz.
The change shown corresponds to a 10\% increase in $|\vS_{\rm tot}|$. 
The solid and dashed curves refer to $\theta_{L}$ and $\theta_{S}$,
respectively.}
\end{center}
\end{figure}

In this section we study the precessional dynamics of double-spin
binaries, with the purpose of building a physical understanding of the
matching performance of single-spin templates observed in Sec.\
\ref{sec2.2}; in particular, we wish to identify what features of
double-spin dynamics, absent in single-spin systems, lead to the low
FFs seen for lower-mass binary configurations. 

From Eqs.~\eqref{S1dot}--\eqref{Lhdot}, we see that the precession of double-spin
binaries preserves both $\mathbf{J}$ and $|\mathbf{L}|$ at timescales
shorter than the radiation-reaction timescale---at which
$|\mathbf{L}|$ decreases steadily. Even at longer timescales, as recognized by
ACST for single-spin binaries and further tested by BCV2 for generic
double-spin binaries, for the vast majority of configurations the direction of the total angular momentum remains almost constant ($\dot{\hJ} \simeq 0$); this behavior is known as \emph{simple precession}.

For single-spin binaries, or for equal-mass binaries if we ignore the spin-spin
interaction, the angle between $\mathbf{L}$ and $\mathbf{S}_{\rm tot}$ 
($\theta_{LS}$) remains fixed all through evolution [according to
Eqs.~\eqref{S1dot}--\eqref{Lhdot}]; for simple precession, this implies that the angle between $\mathbf{L}$ and $\hJ$ ($\theta_L$) must increase, and that the angle between $\mathbf{S}_{\rm tot}$ and $\hJ$ ($\theta_S$) must decrease---both do so monotonically, at the radiation-reaction timescale. In summary, in these binaries the orbital plane precesses while its inclination increases slowly and monotonically.

In Ref.~[\onlinecite{apostolatos2}], Apostolatos investigated the
effect of spin-spin coupling on the dynamical evolution of {\it
equal-mass, equal-spin} BH--BH binaries. He obtained analytical solutions for the
opening-angle products $\hS_{1,2}\cdot\hJ$ and $\hS_1\cdot\hS_2$ to
first order in $S/L$, where $J$ is the total angular momentum, $S$ is
$S_1$ or $S_2$, and $L$ is the orbital angular momentum (if $m_1 \sim
m_2$, then throughout all the inspiral $S \ll L$~\cite{ACST94}).  The
main feature identified by Apostolatos was that the orbital plane not only becomes slowly more inclined (at the radiation-reaction timescale), but the spin-spin interaction also causes a {\it nutation}; namely, an oscillation of the orbital inclination angle ($\theta_L$) at the timescale of the spin-spin interaction.

In the following, we shall relax the
assumption that the BH masses are equal, and we shall investigate what
the consequences are on the evolution of $\hS_{1,2}\cdot\hJ$ and
$\hS_1\cdot\hS_2$.  To simplify our notation we fix $m_1+m_2=1$, and
we introduce the parameter $\delta \equiv m_1-m_2$, which describes
the deviation from the equal-mass case. We keep only terms up to
linear order in $\delta$. We then have (assuming as always $m_1 \geq
m_2$)
\bea
&& m_1 =\frac{1+\delta}{2}\,, \quad m_2=\frac{1-\delta}{2}\,, \\
&& S_1= m_1^2\,\chi_1 \simeq \frac{1}{4}\,(1+2\delta)\,\chi_1\,,\\
&& S_2= m_2^2\,\chi_2 \simeq  \frac{1}{4}\,(1-2\delta)\,\chi_2\,.
\eea
Inserting these definitions into Eqs.~\eqref{S1dot} and \eqref{S2dot} leads to 
\begin{widetext}
\bea
\label{angleevol1}
\frac{d}{dt}(\hS_1\cdot\hJ)&=&
\frac{\omega^2}{2}\,S_2\,\Big [\overbrace{\vphantom{\biggl(}(7-6\delta)}^{\rm spin-orbit} - 
\overbrace{\biggl(1+3\frac{\vJ \cdot \vS_2}{L^2}
\biggr)}^{\rm spin-spin} \Big ]\,\hJ\cdot(\hS_1\times\hS_2)\,, \\
\label{angleevol2}
\frac{d}{dt}(\hS_2\cdot\hJ)&=&
\frac{\omega^2}{2}\,S_1\,\Big [\overbrace{\vphantom{\biggl(}(7+6\delta)}^{\rm spin-orbit} - 
\overbrace{\biggl(1+3\frac{\vJ \cdot \vS_1}{L^2}
\biggr)}^{\rm spin-spin} \Big ]\,\hJ\cdot(\hS_2\times\hS_1)\,, \\
\label{angleevol3}
\frac{d}{dt}(\hS_1\cdot\hS_2)&=& \frac{\omega^2}{2}\,
\Big [\overbrace{\vphantom{\biggl(}(-12\delta)}^{\rm spin-orbit} + 
\overbrace{\biggl(3\frac{\vJ \cdot (\vS_1-\vS_2)}{L^2}\biggr)}^{\rm spin-spin} \Big ]\,\vJ\cdot(\hS_1\times\hS_2)\,.
\eea
\end{widetext}
Following Apostolatos, in deriving these equations we have assumed
that $S \ll L$, $L=J\,[1+\mathcal{O}(S/L)]$, and that the direction of
the total angular momentum remains almost constant during evolution. In Eqs.~\eqref{angleevol1}--\eqref{angleevol3} we
have separated the terms due to spin-orbit and spin-spin interactions.

According to Eq.~\eqref{angleevol3}, in the equal-mass case
($\delta=0$) without spin-spin effects, the angle between $\hS_1$ and
$\hS_2$ is constant; generically, however, the spin-spin and even the
spin-orbit interactions can cause that angle (and hence the
magnitude of $\mathbf{S}_{\rm tot}$) to change.  As first observed by ACST and
further investigated by Apostolatos~\cite{apostolatos2}, this
variation in $|\mathbf{S}_{\rm tot}|$ drives the nutation of the orbital plane: oscillations are superimposed to the monotonic evolution of
the angles between $\hL$ and $\hJ$ ($\theta_L$) and between $\hS_{\rm
tot}$ and $\hJ$ ($\theta_{S}$)~\cite{apostolatos2}, as can be
understood from the following simple argument.  Recall that on
timescales shorter than the radiation-reaction timescale,
$|\mathbf{L}|$ and $\mathbf{J}$ are conserved; using $\delta$ to
denote the change in the dynamical variables incurred during such a
time, we write
\beq
\delta |\mathbf{J}|^2 =0\,, \quad \delta |\mathbf{L}|^2 =0
\eeq
to get
\beq
\label{eqLS}
2\delta (\mathbf{L}\cdot\mathbf{S}_{\rm tot})= -\delta(|\mathbf{S}_{\rm tot}|^2)\,.
\eeq
Using $\delta \mathbf{J}=0$ we then have
\beq
\delta \lambda_L = -\frac{1}{2L J}\delta(|\mathbf{S}_{\rm tot}|^2)
\quad \mathrm{for} \quad
\lambda_{L} = \frac{\mathbf{L}\cdot \mathbf{J}}{L J}\,,
\eeq
and
\beq
\delta \lambda_S =  -\frac{\mathbf{L}\cdot\mathbf{S}_{\rm tot}}
{2 S_{\rm tot}^3 J}\delta(|\mathbf{S}_{\rm tot}|^2)
\quad \mathrm{for} \quad
\lambda_{S} = \frac{\mathbf{S}_{\rm tot}\cdot \mathbf{J}}{S_{\rm tot} J}\,.
\eeq
Thus, when $|\mathbf{S}_{\rm tot}|$ oscillates, the opening angles $\theta_{L}$ and $\theta_{S}$ oscillate as well; in fact, we have 
\bea
\delta{\theta}_{L}&=&\frac{1}{2|\mathbf{L}\times\mathbf{S}_{\rm
tot}|}\delta(|\mathbf{S}_{\rm tot}|^2)\,, \label{dthetaL} \\
\delta{\theta}_{S}&=&\frac{\mathbf{L}\cdot\mathbf{S}_{\rm tot}}
{2|\mathbf{S}_{\rm tot}|^2 |\mathbf{L}\times\mathbf{S}_{\rm tot}|}
\delta(|\mathbf{S}_{\rm tot}|^2) \label{dthetaS}\,, 
\eea 
which suggests that, for the same variation in $|\mathbf{S}_{\rm tot}|^2$,
the nutation is most significant when $\mathbf{L}$ and $\mathbf{S}$
are either nearly aligned or antialigned.  In Fig.~\ref{dopenang},
we plot the relative changes in $\theta_L$ and $\theta_S$ as
functions of the angle between $\mathbf{L}$ and $\mathbf{S}$
($\theta_{LS}=\theta_L+\theta_S$), choosing a fixed positive
$\delta(|\mathbf{S}_{\rm tot}|)$. The change is always positive for
$\theta_{L}$, while it can be negative for $\theta_S$, if
$\theta_{LS}>90^\circ$. In addition, the relative changes {\it
diverge} near $\theta_{LS}\sim 0^\circ$ or $180^\circ$. These features
follow straightforwardly from Eqs.~\eqref{dthetaL} and
\eqref{dthetaS}.

\emph{Spin-spin effects.} When spin-spin effects are included, the
angle between $\hS_1$ and $\hS_2$ oscillates according to the second
term on the right-hand side of Eq.~(\ref{angleevol3}). However, as evidenced by
the FF results for equal-mass binaries given in Sec.\ III B, the amplitude of
these oscillations does not seem to be very large, and the nutation of
the orbital plane does not complicate significantly the waveforms, at least as
evaluated at the leading mass-quadrupole order.

\emph{Spin-orbit effects.} From the first term on the right-hand side of
Eq.~(\ref{angleevol3}), we see that even in the absence of spin-spin
effects, unequal masses (i.e., $\delta \neq 0$) can cause the
evolution of $\hS_1 \cdot \hS_2$, and therefore drive the nutation of the orbital plane.  Spin-orbit effects, which come in at a lower PN order, can
sometimes be more significant than spin-spin effects, especially for
binaries with intermediate mass ratios, such as $(m_1+m_2) = (6+3)
M_\odot$ and $(9+3) M_\odot$ binaries; indeed, these effects could
explain the lower FFs found in Sec.\ III B for those systems.

\begin{figure*}
\begin{center}
\includegraphics[width=6.5in]{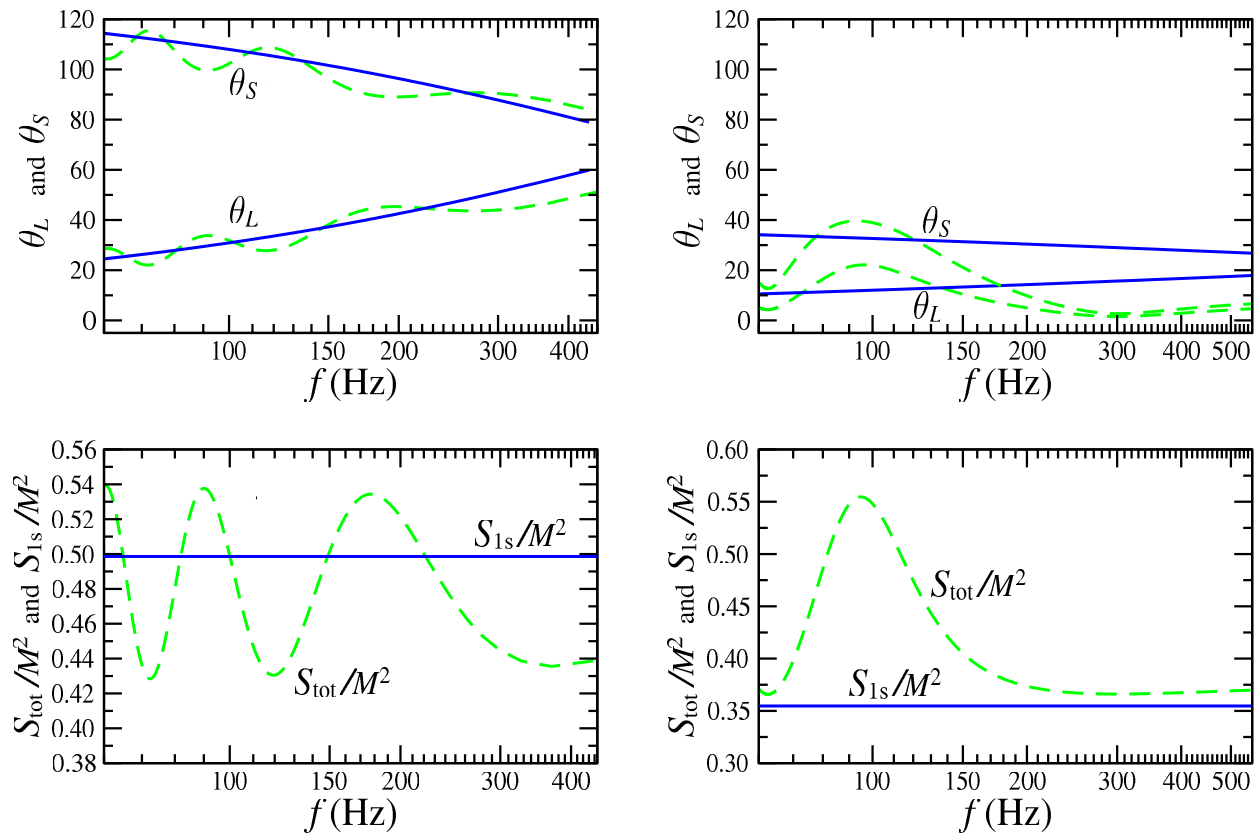}
\caption{\label{lambda} Evolution of the opening angles
$\theta_{L}$ and $\theta_{S}$, and of the total-spin magnitude $S_{\rm tot}$ (all plotted as dashed lines) for double-spin target systems yielding $\mathrm{FF} \ge 0.99$ (left column) and $\mathrm{FF} \simeq 0.94$
(right column) when matched by single-spin templates; the target
system has $(m_1+m_2) = (6+3)M_{\odot}$.  For comparison, the solid
lines show the evolution of the analogous single-spin quantities
[$\theta_{L}$, $\theta_{S}=\arccos(\hS_{1s}\cdot\hJ)$, and $S_{1s}$] 
for the best-fit single-spin systems.}
\end{center}
\end{figure*}

\begin{figure}
\begin{center}
\includegraphics[width=3.2in]{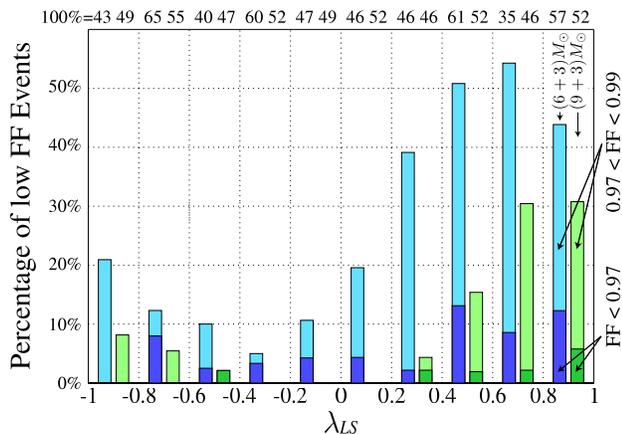}
\caption{\label{lowFFlambda}Percentage of initial spin configurations
that yield FF $\le 0.99$ (light pattern) and 
FF $\le 0.97$ (dark pattern), as a function of the initial opening angle
product $\lambda_{LS} = \hL \cdot \hS_{\rm tot}$, for $(m_1+m_2) =
(6+3)M_{\odot}$ and $(9+3)M_{\odot}$ binaries. The numbers on top show
the total number of configurations (among 500) randomly extracted
within each bin of $\Delta \lambda_{LS} = 0.2$.}
\end{center}
\end{figure}

Examples of combined spin-spin and spin-orbit effects in double-spin
binaries leading to oscillations in $|\vS_{\rm tot}|$, and therefore
$\theta_L$ and $\theta_S$, are shown in Fig.~\ref{lambda} for
systems with $(m_1+m_2) = (6+3) M_\odot$ ($\delta = 1/3$),
and with initial local parameters such that $\mathrm{FF} \ge 0.99$ (on
the left) and $\mathrm{FF} \sim 0.94$ (on the right).  The nutation
behavior evident in these figures is well described by the
approximated equations \eqref{dthetaL} and \eqref{dthetaS}.  For comparison
we show (as continuous lines) also the evolution of the analogous
quantities in the best-fit single-spin configurations; for
these, the opening angles $\theta_L$ and $\theta_S$ evolve
monotonically. Lower overlaps seem to correspond to initial conditions
for which nutation is rather significant and overwhelms the
underlying monotonic evolution. In Fig.\ \ref{lowFFlambda} we show the
percentage of configurations with FF $\le 0.99$ (light pattern) 
and FF $\le 0.97$ (dark pattern) as a function of the initial $\lambda_{LS} \equiv \hLN \cdot \hat{\mathbf{S}}_{\rm tot}$ (evaluated at the starting frequency \cite{notels}), for $ (6+3) M_\odot$ and $(9+3) M_\odot$ binaries.  The plot suggests that lower FFs are more likely to occur when the initial $\hL \cdot \hS_{\mathrm{tot}} \simeq \pm 1$,
which is consistent with the fact that nutation is most
significant when $\theta_{LS}\sim 0^\circ$ or $180^\circ$~\cite{notespin}.

Figure \ref{lowFFlambda} shows also an asymmetry in the distribution of low FFs,
which are denser when $\theta_{LS} < 90^\circ$. Currently we do not have a clear understanding of this behaviour \cite{noteperhaps}. It is worth pointing out that this asymmetry and, more in general, the low FFs observed could be due to other features of double-spin dynamics that cannot be reproduced by single-spin systems,
but that are difficult to dig out by analysis or numerical experiment,
in the absence of a full analytical solution to the precession
equations. Moreover, some of the low FFs might be due
to shortcomings in our numerical optimization procedure in cases where
the match surface in the search parameter space has an especially
convoluted geometry.

\section{Template space and number of templates}
\label{seccounting}

In this section we estimate the number of single-spin templates
necessary to search for double-spin signals with single masses in the
$[3,15]M_\odot$ range. To do this, we compute the average projected
metric $\overline{g^\mathrm{proj}_{ij}}$ in the $(M_s,\eta_s)$ region
delimited by the dashed and dotted contour of Fig.\ \ref{fig:scatter},
following the procedure described in Sec.\ VI of Ref.\
\cite{pbcv1}. We notice the presence of reduction curves connecting
(roughly) the segment $\chi_{1s} \in [0,2], \kappa_{1s} = 0$ to the
entire $(\chi_{1s},\kappa_{1s})$ plane in the search template space.
Thus, we select a 3-D reduced template space corresponding to
$(M_s,\eta_s)$ within the quadrilateral with vertexes $(15 M_\odot,
0.08)$, $(3.25 M_\odot, 0.275)$, $(32.5 M_\odot, 0.23)$, $(24.5
M_\odot, 0.385)$, to $\chi_{1s} \in [0,2]$, and to $\kappa_{1s} = 0$.
Additional subfamilies might be needed to deal with certain
singularities that we observe in the reduction curves as $\kappa_{1s}$
gets close to $\pm 1$, but our selection should already give us an
acceptable idea of the number of necessary templates, which is
computed according to
\begin{equation}
{\cal N}_\mathrm{templates}= \frac{ {\displaystyle\int} \!
\sqrt{\left|\det \overline{g^\mathrm{proj}_{i'j'}}\right|} dM_s \,
d\eta_s \, d\chi_{1s}}{ \left[2 \sqrt{(1-{\rm MM})/3}\right]^3},
\end{equation}
where the primed indices $i'$, $j'$ run through $M_s$, $\eta_s$, and
$\chi_{1s}$; the metric is averaged over 1,000 sets of target extrinsic
parameters. The integral is carried out by evaluating the projected
metric (a computationally expensive operation) at 80 points within the
integration region, and filling it by \emph{natural neighbor
interpolation} \cite{nninterp}. The final result is
$\mathcal{N}_\mathrm{templates} \simeq$ 320,000 for $\mathrm{MM} =
0.98$ (not including a reduction mismatch of $\sim 0.01$ incurred along the reduction curves \cite{pbcv1}). Given the uncertainties implicit in the numerical computation of the metric, in the interpolation, in the choice of the reduction curves, and in the actual placement of the templates in the bank, this number should be
understood only as an order-of-magnitude estimate. Most of the
templates, by a factor of many, come from the lower part of the
integration region (i.e., from the lowest $\eta_s$ for any given
$M_s$); about 100,000 out of 320,000 come from the region with $\chi_{1s} < 1$.

\section{Estimation of binary parameters}
\label{sec4}

Since the single-spin template family contains only a subset (of lower dimensionality) of all possible double-spin waveforms, we cannot expect to obtain estimates of all the physical parameters of double-spin systems from a single-spin--template search. The most straightforward way to recover those parameters would be to perform a search using double-spin templates after
single-spin templates have yielded a detection; such a follow-on search may be computationally feasible, since double-spin templates will then be applied only to the data stretches that have been established to contain signals.
However, it is still meaningful to analyze the parameter-estimation performance of the single-spin template family, since any constraints on source parameters will decrease the size of the double-spin template bank necessary for the follow-on search, lowering the computational cost even further. In addition, this analysis can offer further useful insight into the FF map of double-spin into single-spin waveforms. 

Double-spin target waveforms are parametrized by twelve parameters (seven intrinsic, and five extrinsic), while single-spin search templates are parametrized by nine (four intrinsic, $M_{s}$, $\eta_{s}$, $\chi_{1s}$, and $\kappa_{1s}$; and five extrinsic). Thus, maximizing the match over the search parameters induces a map from the $12$-parameter target-signal space into the
$9$-parameter template space. The inverse map takes each point in the template space into a 3-dimensional manifold in the target-signal space, whose size indicates the extent to which the physical parameters of the double-spin binary can be constrained. Unfortunately, evaluating the size of the inverse image requires computational resources well beyond what is currently available to us.
Furthermore, such a procedure can only be meaningful after 
\emph{statistical error} has also been taken into account: because the detector output contains also noise, the template parameters at which we obtain the
maximal correlation between template and data (the \emph{actual} projection point) will differ, by a random statistical error, from the parameters
at which the correlation between template and \emph{signal} is highest (the {\it theoretical} projection point) \cite{note05}. We leave the quantitative study of statistical errors to a forthcoming paper~\cite{pbcv3}.

In this paper we take a semi-quantitative approach. We ignore the extrinsic parameters, so we study the map of seven parameters into four; then, we identify a number of intrinsic parameters of the double-spin binary (our \emph{target observables}), and we explore how well we can estimate their values using functions of the four intrinsic parameters of the best-fit single-spin template (our \emph{estimators}). We use three general criteria in the choice of target observables and estimators:
\begin{description}
\item[Consistency.] The target observables and estimators should coincide when the double-spin system is dynamically equivalent to a single-spin system (according to the criteria spelled out in Sec.\ I).
\item[Robustness.] The definitions of the target observables and of the estimators should be independent of the detector noise curve; equivalently, the target observables should be (almost) conserved quantities (we already know that the four template parameters, and hence the estimators, are conserved).
\item[Strong influence.] The target observables, and therefore the estimators, should have a strong influence to the waveforms. Quantitatively, we can require the mismatch metric to have large components along the direction of change of the target observables and estimators; it follows that the target observables and estimators should remain essentially constant along eventual reduction curves.

This criterion is important for two reasons: first, the 7-to-4-dimensional FF map in unlikely to preserve unessential features of the target space; second, even if the map preserved these features, statistical error would inevitably spoil their estimation, because the associated mismatch-metric components are small. A third reason, contingent on our implementation of the FF search, is that we stop the maximization of the match whenever this reaches 0.99; this adds a dominant artificial fluctuation (roughly corresponding to the statistical error for a S/N of 100) to parameter estimation.
\end{description}
Since our search template family possesses a family of approximate reduction curves, it will be generally possible to estimate efficiently only three independent target observables, and four only for very high S/N. Our hope is then to find three target-parameter--estimators pairs that satisfy all three criteria, and four pairs that satisfy the consistency and robustness criteria.
{\renewcommand{\arraystretch}{1.5}
\begin{table*}
\begin{tabular}{c|c||r|r|r|r|r|r}
 &  & $(6+3)\,M_\odot$   & $(9+3)\,M_\odot$   & $(12+3)\,M_\odot$   & $(10+10)\,M_\odot$   & $(15+10)\,M_\odot$   & $(15+15)\,M_\odot$  \\  &  & ($500$ points)   & ($500$ points)   & ($100$ points)   & ($500$ points)   & ($100$ points)   & ($100$ points)  \\ \hline \hline 
$M$ & $(\overline{M}_{s}-M)/M$ &  $+ 0.0232$ &  $+ 0.0021$ &  $-0.0066$ &  $-0.0351$ &  $-0.0206$ &  $-0.0579$\\ 
 & $\Delta M_{s}/M$ & $ 0.0817$ & $ 0.0755$ & $ 0.0631$ & $ 0.0437$ & $ 0.0603$ & $ 0.0639$\\ 
 & $1$-$\sigma$/$3$-$\sigma$ percentage & $ 75.6\%$/$ 98.0\%$ & $ 84.0\%$/$ 97.4\%$ & $ 79.0\%$/$ 97.0\%$ & $ 76.4\%$/$ 98.6\%$ & $ 70.0\%$/$ 99.0\%$ & $ 72.0\%$/$ 100.0\%$\\ 
 \hline$\eta$ & $\overline{\eta}_{s}-\eta$ &  $-0.0057$ &  $+ 0.0022$ &  $+ 0.0038$ &  $+ 0.0191$ &  $+ 0.0122$ &  $+ 0.0362$\\ 
 & $\Delta\eta_{s}$ & $ 0.0268$ & $ 0.0254$ & $ 0.0178$ & $ 0.0219$ & $ 0.0283$ & $ 0.0387$\\ 
 & $1$-$\sigma$/$3$-$\sigma$ percentage & $ 72.6\%$/$ 99.0\%$ & $ 84.4\%$/$ 98.8\%$ & $ 81.0\%$/$ 98.0\%$ & $ 76.6\%$/$ 98.6\%$ & $ 74.0\%$/$ 99.0\%$ & $ 74.0\%$/$ 100.0\%$\\ 
 \hline$\mathcal{M}$ & $(\overline{\mathcal{M}}_{s}-\mathcal{M})/\mathcal{M}$ &  $-0.0004$ &  $+ 0.0015$ &  $+ 0.0021$ &  $+ 0.0055$ &  $+ 0.0033$ &  $+ 0.0142$\\ 
 & $\Delta\mathcal{M}_{s}/\mathcal{M}$ & $ 0.0074$ & $ 0.0106$ & $ 0.0104$ & $ 0.0092$ & $ 0.0145$ & $ 0.0192$\\ 
 & $1$-$\sigma$/$3$-$\sigma$ percentage & $ 71.6\%$/$ 99.0\%$ & $ 69.6\%$/$ 99.2\%$ & $ 66.0\%$/$ 100.0\%$ & $ 71.2\%$/$ 99.2\%$ & $ 70.0\%$/$ 98.0\%$ & $ 80.0\%$/$ 99.0\%$\\ 
 \hline\hline$\vphantom{\Bigg|}\frac{\mathbf{S}_{\rm eff}\cdot\hat{\mathbf{L}}_{N}}{M^2}$ & $\overline{ \left(\frac{ \hLN \cdot \mathbf{S}_{\rm eff} }{M^2}\right)_{\!s}-\frac{ \hLN \cdot \mathbf{S}_{\rm eff} }{M^2}}$ &  $+ 0.0236$ &  $+ 0.0210$ &  $+ 0.0129$ &  $-0.0071$ &  $-0.0015$ &  $+ 0.0004$\\ 
 & $\Delta \left[\left(\frac{ \hLN \cdot \mathbf{S}_{\rm eff} }{M^2}\right)_{\!s}-\frac{ \hLN \cdot \mathbf{S}_{\rm eff} }{M^2}\right]$ & $ 0.1290$ & $ 0.1130$ & $ 0.1010$ & $ 0.0747$ & $ 0.0886$ & $ 0.0917$\\ 
 & $1$-$\sigma$/$3$-$\sigma$ percentage & $ 79.4\%$/$ 98.0\%$ & $ 73.4\%$/$ 98.8\%$ & $ 68.0\%$/$ 100.0\%$ & $ 75.8\%$/$ 99.0\%$ & $ 66.0\%$/$ 100.0\%$ & $ 73.0\%$/$ 98.0\%$\\ 
 \hline\hline$\vphantom{\Bigg|}\frac{\mathbf{S}_{\rm eff}}{M}$ & $\frac{1}{M}\overline{\left[\left(\frac{\mathbf{S}_{\rm eff}}{M}\right)_{\!s}-\frac{\mathbf{S}_{\rm eff}}{M}\right]}$ &  $+ 0.0253$ &  $-0.0264$ &  $-0.0183$ &  $+ 0.0375$ &  $+ 0.0028$ &  $+ 0.0387$\\ 
 & $ \frac{1}{M} \Delta \left[\left(\frac{\mathbf{S}_{\rm eff}}{M}\right)_{\!s}-\frac{\mathbf{S}_{\rm eff}}{M}\right]$ & $ 0.1920$ & $ 0.1480$ & $ 0.1050$ & $ 0.1170$ & $ 0.1860$ & $ 0.1460$\\ 
 & $1$-$\sigma$/$3$-$\sigma$ percentage & $ 75.0\%$/$ 98.6\%$ & $ 72.6\%$/$ 98.6\%$ & $ 67.0\%$/$ 99.0\%$ & $ 75.0\%$/$ 98.8\%$ & $ 74.0\%$/$ 100.0\%$ & $ 79.0\%$/$ 99.0\%$\\ 
 \hline$\chi_{\rm tot}$ & $\overline{(\chi_{\rm tot})_s-\chi_{\rm tot}}$ &  $+ 0.0627$ &  $-0.0053$ &  $-0.0140$ &  $+ 0.0174$ &  $+ 0.0036$ &  $+ 0.0149$\\ 
 & $\Delta\left[{(\chi_{\rm tot})_s-\chi_{\rm tot}}\right]$ & $ 0.2040$ & $ 0.1610$ & $ 0.1190$ & $ 0.0901$ & $ 0.1450$ & $ 0.1220$\\ 
 & $1$-$\sigma$/$3$-$\sigma$ percentage & $ 81.6\%$/$ 98.2\%$ & $ 81.8\%$/$ 97.8\%$ & $ 85.0\%$/$ 97.0\%$ & $ 78.8\%$/$ 98.0\%$ & $ 71.0\%$/$ 99.0\%$ & $ 80.0\%$/$ 98.0\%$
\end{tabular}
\caption{Systematic biases, rms deviations, and percentage of samples within $\pm 1$ and 3 deviations of the average, for six target-observable--estimator pairs. The mass configurations are those studied in Sec.\ \ref{sec2.2}.
\label{tab:est}}
\end{table*}}

\begin{figure*}
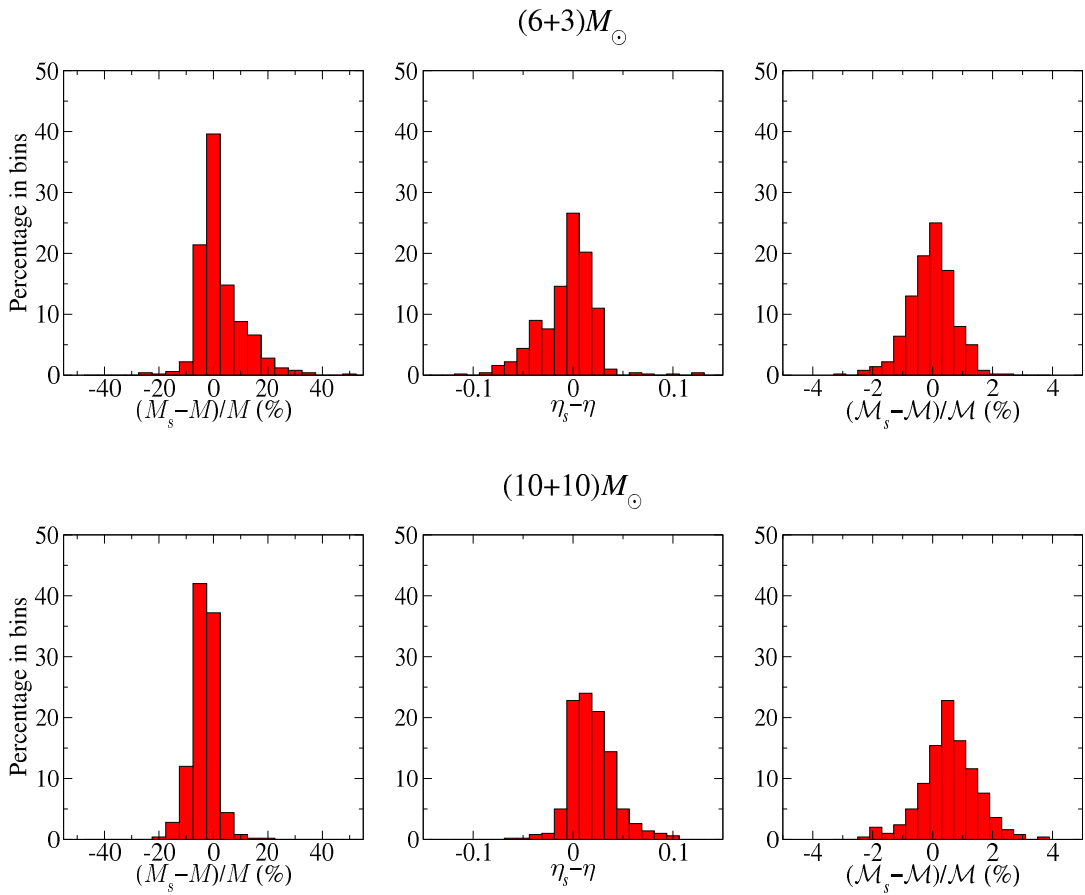

\begin{center}
\begin{tabular}{c}
\includegraphics[width=0.8\textwidth]{M-6-3.eps} \\
\vspace{0.cm}
\\
\includegraphics[width=0.8\textwidth]{M-10-10.eps}
\end{tabular}
\caption{\label{fig:masshist} Distribution of errors for the target
observables $M$, $\eta$, and $\mathcal{M}$, as estimated by $M_{s}$,
$\eta_{s}$ and $\mathcal{M}_{s}$, for 500 double-spin binaries with
$(m_1+m_2) = (6+3)M_\odot$ and $(10+10)M_\odot$, maximal spins, and
uniform distributions of local parameters. The $\mathcal{M}_s$
distribution has the smallest bias and dispersion; the $M_s$ and
$\eta_s$ distributions have much larger dispersion and are skewed in
opposite directions (as needed to reduce the dispersion of
$\mathcal{M}_s \equiv M_s \eta_s^{3/5}$).}
\end{center}
\end{figure*}

It is straightforward to see that using $M_{s}$ to estimate $M$ and $\eta_s$ to estimate $\eta$ (or in shorthand, $M_{s} \rightarrow M $ and $\eta_{s}
\rightarrow \eta$) satisfies the consistency and robustness criteria. We shall pay a special attention to the estimation of chirp mass according to $\mathcal{M}_{s} \equiv M_{s} \eta_{s}^{3/5} \rightarrow \mathcal{M} \equiv M \eta^{3/5}$, which satisfies all three criteria (in particular, chirp mass is conserved very accurately along reduction curves). In upper part of Table~\ref{tab:est} we characterize the distribution of estimation error for $M$, $\eta$, and $\mathcal{M}$, for the same systems (with fixed masses and random local parameters) used in Sec.\ \ref{sec2.2} to compute FFs. Each section shows the estimation bias (defined as the average of the error, and measuring a systematic displacement between observables and estimators that can in principle be removed), its rms deviation (measuring the intrinsic uncertainty in the estimation), and the percentage of estimators enclosed within 1-deviation and 3-deviation intervals (measuring the normality of the distribution).
The chirp mass, which satisfies all three criteria, is 
indeed estimated with higher relative accuracy than both $M$ and
$\eta$. Even after statistical errors are taken into account, this accuracy should be retained for $\mathcal{M}$ better than for $M$ and $\eta$. The distributions of $M_s$, $\eta_s$, and $\mathcal{M}_s$ are also histogrammed in 
Fig.~\ref{fig:masshist}, for $(m_1+m_2) = (6+3)M_\odot$ and $(10+10)M_\odot$ binaries. 

\begin{figure*}
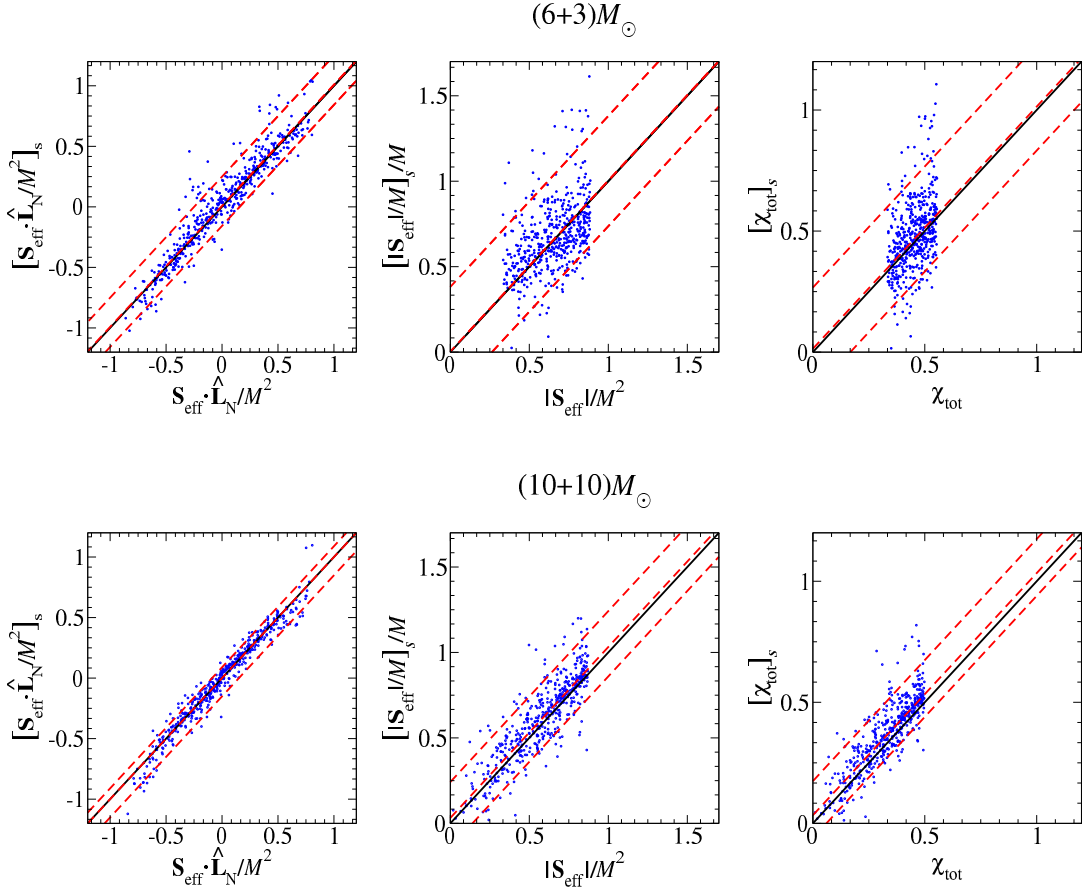

\begin{center}
\begin{tabular}{c}
\includegraphics[width=0.8\textwidth]{LS-6-3.eps} \\
\vspace{0.2cm}\\
\includegraphics[width=0.8\textwidth]{LS-10-10.eps}
\end{tabular}
\caption{\label{fig:histospin} Estimation of spin-related parameters
 for 500 double-spin binaries with $(m_1+m_2) = (6+3)M_\odot$ and $(10+10)M_\odot$, maximal spins, and uniform distributions of local parameters. The three columns display the correlations between the (target-observable, estimator) pairs $(\mathbf{S}_{\rm eff}\cdot \hLN/M^2,[\mathbf{S}_{\rm eff}\cdot \hLN/M^2]_s)$, $(|{\bf S}_{\rm eff}|/M^2,[|{\bf S}_{\rm eff}|/M]_s/M)$, and $(\chi_{\rm tot},[\chi_{\rm tot}]_s)$. In each panel, three dashed lines are used to indicate the median of the bias and the 90\% percentiles above and below it.  The solid line marks the line of zero bias. Lower biases correspond to solid lines closer to the central dashed lines; lower dispersions correspond to closer outer dashed lines.}
\end{center}
\end{figure*}

It is very hard to identify additional target-parameter--estimator pairs that
satisfy all three criteria, mainly because double-spin binaries lack
conserved quantities that clearly dominate the waveforms; so the two additional target observables that can be estimated efficiently may not have simple physical meanings.  When spin-spin effects are negligible, the only truly conserved quantity that could be interesting for our purposes is $\mathbf{S}_{\rm eff} \cdot \hLN$, with
\beq
\mathbf{S}_{\rm eff}\equiv
\left(1+\frac{3m_2}{4m_1}\right)\mathbf{S}_1+\left(1+\frac{3m_1}{4m_2}\right)\mathbf{S}_2\,.
\eeq
[The magnitudes of the individual spins
do not satisfy the consistency criterion, since single-spin binaries (e.g., with $m_1$ spinning) require $|\mathbf{S}_{1s}|\rightarrow |\mathbf{S}_{1}|$, while double-spin, equal-mass binaries require
$|\mathbf{S}_{1s}| \rightarrow |\mathbf{S}_1+\mathbf{S}_2|$.] Therefore, we choose the target observable $\mathbf{S}_{\rm eff}\cdot \hLN/M^2$, which is conserved and closely related to the opening angle between $\mathbf{S}_{\rm eff}$ and $\hLN$, and hence to the depth of the modulation caused by orbital precession. The estimator is naturally
\begin{equation}
\label{estSeff}
\left[\frac{\mathbf{S}_{\rm eff}\cdot \hLN}{M^2}\right]_{s}
\equiv \left(1+\frac{3m_{2s}}{4m_{1s}}\right)\frac{m_{1s}^2}{M_s^2}\chi_{1 s}\kappa_{1 s}\,,
\end{equation}
where $(m_{1,2})_{s}\equiv {\rm Re}\left[(1 \pm \sqrt{1-4\eta_{s}})/2\right]M_{s}$ (taking the real part becomes necessary when $\eta_{s}>0.25$).
As we see from Table~\ref{tab:est}, this observable can be estimated with bias within $[-0.01,+0.02]$, and with rms deviation $0.07$--$0.13$. Although
conserved, this observable is not quite constant along reduction
curves.
In the left panels of Fig.~\ref{fig:histospin}, we plot the distribution of the pairs $(\mathbf{S}_{\rm eff}\cdot \hLN/M^2,[\mathbf{S}_{\rm eff}\cdot\hLN/M^2]_s)$ for $(m_1+m_2) = (6+3)M_\odot$ and $(10+10)M_\odot$
binaries. In each case we conclude that the target observable and the estimator are strongly correlated; however, the dispersion is noticeably smaller for $(10+10)M_\odot$ than for $(6+3)M_\odot$ binaries; this must be because for equal-mass binaries only spin-spin effects can cause differences between double-spin and single-spin waveforms.

In the light of Eq.~\eqref{Lhdot}, we choose the third target
observable as $|\mathbf{S}_{\rm eff}|/M$, which measures the
instantaneous angular precession frequency divided by $\omega^2$. This
quantity is conserved only for single-spin binaries and for equal-mass binaries with negligible spin-spin effects, so it does not completely satisfy the robustness criterion. In our study, we use the value of $|\mathbf{S}_{\rm eff}|/M$ at the initial frequency (40 or 60 Hz) from which the equations of motion are integrated. [However, it would be more reasonable to evaluate it at the frequency at which the detector is most sensitive, or to weight its values at different frequencies according to the detector noise spectrum.] The estimator is
\beq
\left[\frac{|\mathbf{S}_{\rm eff}|}{M}\right]_{s} = 
\left(1+\frac{3m_{2s}}{4m_{1s}}\right)\frac{m_{1s}^2}{M_s} \chi_{1 s}
\,. 
\eeq
To make the observable and the estimator dimensionless, we divide both by $M$; Table~\ref{tab:est} then shows that $|\mathbf{S}_{\rm eff}|/M^2$ can be estimated with bias within $[-0.03,+0.04]$, and rms deviation $0.11$--$0.19$.
Despite the apparent physical meaning of $\left[|\mathbf{S}_{\rm eff}|/M\right]_{s}$, this observable is also not conserved well along reduction curves. Thus, we might as well use a more familiar target observable, $\chi_{\rm tot} \equiv \mathbf{S}_{\rm tot}/M_s^2$, as estimated by 
$[\chi_{\rm tot}]_s=\chi_{1 s}m_{1s}^2/M^2$. This pair satisfies the consistency criterion, and it changes through the inspiral at a level similar to $\mathbf{S}_{\rm eff}/M^2$. Table~\ref{tab:est} shows that $\chi_{\rm tot}$ can be estimated with bias within $[-0.005, +0.06]$, and rms deviation $0.10$--$0.20$. 
In the center and right panels of Fig.~\ref{fig:histospin}, we plot $({\rm target},{\rm estimator})$ distributions for these two spin observables, again for for $(m_1+m_2) = (6+3)M_\odot$ and $(10+10)M_\odot$ binaries. It is clear from the plots that the accuracy of estimation is poorer than for
$\mathbf{S}_{\rm eff}\cdot \hLN/M^2 \rightarrow [\mathbf{S}_{\rm eff}\cdot  \hLN/M^2]_s$; again, the dispersion is smaller for the $(10+10)M_\odot$ binaries. For $(6+3)M_\odot$ binaries, the worse accuracy can be attributed in part to the fact that these two target observables are not as well conserved as
$\mathbf{S}_{\rm eff}\cdot \hLN/M^2$ during evolution. 

\section{Conclusions}
\label{sec5}

As originally pointed out by ACST~\cite{ACST94}, the dynamics of double-spin precessing binaries become equivalent to the dynamics of single-spin binaries (at least for the purpose of computing gravitational waveforms at the leading mass-quadrupole order) in two limits: 
equal masses, when spin-spin effects can be neglected (then $\mathbf{S}_1 \rightarrow \mathbf{S}_\mathrm{tot}$), and very different masses (then $\mathbf{S}_1$ tends to the spin of the heavier body).
Building on this observation, on the results of Refs.~\cite{K,apostolatos2,bcv2}, and on the (justified) assumption that spin-spin effects contribute mildly to the PN binding energy and GW flux of the binary for mass configurations of interest to ground-based GW interferometers, we conjectured that single-spin templates (as defined in Sec.\ \ref{sec2.2}) can be used effectually to search for double-spin precessing binaries with such masses.

We tested our conjecture by evaluating the FF between the single-spin and double-spin families, and we found confirmation in the very high FF values [see Table~\ref{Tab1} and Fig.~\ref{myhist}] for 
equal-mass binaries of both low and high total masses. FFs were high also for unequal-mass binaries, except for few initial spin configurations. As discussed in Sec.~IIID, for those configurations the evolution of the opening angles between $\hat{\mathbf{J}}$ and $\hat{\mathbf{L}}_N$ and between $\hat{\mathbf{J}}$ and $\hat{\mathbf{S}}_{1,2}$ seem to contain large oscillations, induced by spin-spin and non-equal--mass effects, that cannot be reproduced sufficiently well by single-spin systems.

The region in the single-spin parameter space needed to match
double-spin binaries with $(m_1,m_2) = [3,15]M_\odot \times
[3,15]M_\odot$ is shown in Fig.~\ref{fig:scatter}. Using the LIGO-I
design sensitivity, we counted (very roughly) as $\sim 320,000$ the
number of templates required to yield a minimum match of 0.97. The
number of BCV2 templates needed for a similar mass range is somewhat larger. More generally, with
respect to the detection template families introduced in
Refs.~\cite{apostolatos1,apostolatos2,GKV,bcv2,GK}, the advantage of
the quasi-physical single-spin family suggested in this paper is the
possibility of estimating the parameters of the source. In
Sec.~\ref{sec4} we computed the systematic errors that would affect
the measurement: the total mass $M$ could be estimated with a
fractional bias within $[-6\%,+3\%]$ and a fractional rms deviation of
$5\%$--$8\%$; the symmetric mass ratio $\eta$ could be estimated with
a bias within $[-0.06,+0.04]$ and an rms deviation of $0.02$--$0.04$;
the chirp mass ${\cal M}$ could be estimated with a fractional bias
within $[-0.04\%,+0.01\%]$ and a fractional rms deviation of
$0.7\%$--$2\%$.  We also proposed estimators for certain spin
parameters of the double-spin system. For example, the parameter
$(\mathbf{S}_{\rm eff}\cdot \hLN)/{M^2}$ [where $\mathbf{S}_{\rm eff}$
is defined by Eq.~(\ref{estSeff})], which is conserved when spin-spin
effects can be neglected, could be estimated with a bias within
$[-0.01,+0.03]$ and an rms deviation of $0.07$--$0.13$; the parameter
$\chi_{\rm tot}\equiv\mathbf{S}_{\rm tot}/M^2$ could be estimated with
a bias within $[-0.005, +0.06]$ and an rms deviation of
$0.10$--$0.20$. However, since the mismatch metric has small
components along the directions of these spin estimators, we expect
that (at least for moderate S/N) statistical errors will always be
dominant over the systematic errors discussed here. We defer the study
of statistical errors to a forthcoming paper~\cite{pbcv3}.

In evaluating the performance of the quasi-physical single-spin template family, we have assumed a \emph{uniform} distributions for the initial local parameters (spin and orientation angles) of the double-spin target model. It would be interesting in the future to redo our analyses assuming more realistic nonuniform distributions derived from astrophysical considerations. The only available results for spin distributions in BH--BH binaries (unfortunately, with a single spin) and in NS--BH binaries were obtained using population-synthesis techniques~\cite{VK,Gpc}. For the case of binaries formed in globular clusters, there is no theoretical argument to suggest any particular spin distribution.

Last, recent studies of PN spin-spin effects~\cite{JS} suggest that, for binaries with comparable masses, the two BH spins may have become roughly \emph{locked} into a fixed relative configuration by the time the GWs enter the band of good interferometer sensitivity. If these results are confirmed, they could provide preferred initial spin conditions, and by reducing the variability of GW signals, they could 
help to explain the good performance of our single-spin template family. We must however note that we were motivated in proposing our single-spin templates by the assumption that spin-spin effects never dominate, while for locking to occur spin-spin effects seem to be crucial~\cite{JS}. 

\acknowledgments The research of Y.C.\ and Y.P.\ was supported by
NSF grant PHY-0099568 and NASA grant NAG5-12834. Y.C.\ is also
supported by the David and Barbara Groce Fund at the San Diego
Foundation.  M.V.\ is grateful to the \emph{Institut d'Astrophysique}
in Paris for hospitality during the completion of this work; M.V.'s
research was supported by the LISA Mission Science Office at the Jet
Propulsion Laboratory, Caltech, where it was performed under contract
with the National Aeronautics and Space Administration.


\begin{thebibliography}{99}
\bibitem{note01} In fact, such ``templates'' are not complete waveforms, but contain all the ingredients needed to construct the waveforms for any choice of the extrinsic parameters. Thus, they are perhaps better described as ``multitemplates,'' or ``pseudotemplates.''
\bibitem{note02} One example is the
initial phase $\Phi_0$ in nonspinning-binary templates: fixing all the
other parameters, the value of $\rho$ for any $\Phi_0$ (and indeed,
the maximum $\rho$ over all $\Phi_0$) is obtained by combining
$\rho(\Phi_0 = 0)$ and $\rho(\Phi_0 = \pi/2)$.
\bibitem{note04} To be fully consistent in using Eq.\ \eqref{omegadot}
up to 3.5PN order, we should include high-order spin-orbit and
spin-spin terms that are not available at this time.
\bibitem{notels} For double-spin binaries $\lambda_{LS}$ is not constant, since the evolution of $|\mathbf{S}_{\rm tot}|$ can lead it to oscillate [see Eq.~\eqref{eqLS}]; however, these oscillations do not destroy the qualitative picture see in Fig.~\ref{lowFFlambda}.
\bibitem{notespin} When $\hL$ and $\hS_{\mathrm{tot}}$ are almost aligned or anti-aligned, the FF is again high, since the total waveform modulation induced by orbital procession is negligible. This effect competes with nutation in determining the eventual FF, but it dominates only for a small range of $\theta_{LS}$.
\bibitem{noteperhaps} One conjecture is the following: 
in the single-spin case, $\theta_L$ always 
increases while $\theta_S$ decreases in time. In the double-spin 
case, oscillations are superimposed; if $\theta_{LS} > 90^\circ$, 
$\theta_L$ increases, and $\theta_S$ decreases (or viceversa); while  
if $\theta_{LS} < 90^\circ$, $\theta_L$ and $\theta_S$ 
either both increase or decrease in time. This difference might be
the origin of the poor performance of single-spin templates $\theta_{LS} < 90^\circ$.
\bibitem{note05} Qualitatively, for high-S/N events, the actual projection point lies within the template-space contour that has mismatch $\sim (\mathrm{S}/\mathrm{N})^{-1}$ with the theoretical projection point.
\end{thebibliography}
\end{document}